\documentclass{jfm}
\usepackage{graphicx}
\usepackage{epstopdf,epsfig}
\usepackage{amsmath}
\usepackage{hyperref}
\hypersetup{hidelinks}
\usepackage{booktabs}
\usepackage[dvipsnames]{xcolor}
\usepackage{siunitx}

\def\symbfit{\boldsymbol}

\usepackage[T1]{fontenc}
\usepackage{stix2}

\def\bu{\symbfit{u}}
\def\del{\symbfit{\nabla}}
\def\bx{\symbfit{x}}
\def\dV{\,\mathrm{d}V}
\def\dA{\,\mathrm{d}A}
\def\pd{\partial}
\def\eps{\varepsilon}
\def\J{\mathcal{J}}
\def\K{\mathcal{K}}
\def\P{\mathcal{P}}

\graphicspath{{Figs/}}

\shorttitle{Mixing and APE in stratified flows}
\shortauthor{C. J. Howland, J. R. Taylor and C. P. Caulfield}

\title{Quantifying mixing and available potential energy in vertically periodic simulations of stratified flows}

\author{
    Christopher J. Howland\aff{1} \corresp{\email{\href{mailto:c.j.howland@outlook.com}{c.j.howland@outlook.com}}},
    John R. Taylor\aff{2}
    \and C. P. Caulfield\aff{3,2}
    }

\affiliation{
    \aff{1}Physics of Fluids Group, Max Planck Center for Complex Fluid Dynamics, MESA+ Institute and J.M. Burgers Centre for Fluid Dynamics, University of Twente, P.O. Box 217, 7500AE Enschede, Netherlands
    \aff{2}Department of Applied Mathematics and Theoretical Physics, Centre for Mathematical Sciences, University of Cambridge, Wilberforce Road, Cambridge CB3 0WA, UK
    \aff{3}BP Institute, University of Cambridge, Madingley Road, Cambridge CB3 0EZ, UK
    }

\begin{document}

\maketitle

\begin{abstract}
    Turbulent mixing exerts a significant influence on many physical processes in the ocean.
    In a stably stratified Boussinesq fluid, this irreversible mixing describes the conversion of available potential energy (APE) to background potential energy (BPE).
    In some settings the APE framework is difficult to apply and approximate measures are used to estimate irreversible mixing.
    For example, numerical simulations of stratified turbulence often use triply periodic domains to increase computational efficiency.
    In this setup however, BPE is not uniquely defined and the method of \cite{winters_available_1995} cannot be directly applied to calculate the APE.
    We propose a new technique to calculate APE in periodic domains with a mean stratification.
    By defining a control volume bounded by surfaces of constant buoyancy, we can construct an appropriate background buoyancy profile $b_\ast(z,t)$ and accurately quantify diapycnal mixing in such systems.
    This technique also permits the accurate calculation of a finite amplitude local APE density in periodic domains.
    The evolution of APE is analysed in various turbulent stratified flow simulations.
    We show that the mean dissipation rate of buoyancy variance $\chi$ provides a good approximation to the mean diapycnal mixing rate, even in flows with significant variations in local stratification.
    When quantifying measures of mixing efficiency in transient flows, we find significant variation depending on whether laminar diffusion of a mean flow is included in the kinetic energy dissipation rate.
    We discuss how best to interpret these results in the context of quantifying diapycnal diffusivity in real oceanographic flows.
\end{abstract}


\vspace{-6ex}
    
\section{Introduction}

The transport of heat and salt across surfaces of constant density (isopycnals) in the ocean provides a vital contribution to the closure of the ocean's energy budget \citep{wunsch_vertical_2004,hughes_available_2009}.
As originally highlighted by \cite{munk_abyssal_1966}, such a diapycnal flux arising from molecular diffusion alone is insufficient to balance the generation of dense water in polar regions and close the global circulation.
Turbulence therefore plays an important role in enhancing mixing through the stirring of tracer fields (such as temperature or salinity) and the subsequent generation of small-scale gradients.
In the ocean interior, turbulence is often associated with breaking internal waves \citep{mackinnon_climate_2017}, which in turn lead to mixing that is strongly intermittent in both space and time.
Identifying the mechanisms by which turbulence is generated, and how much mixing can be associated with them, is vital in understanding and accurately modelling the transport of tracers through the ocean.

Here we define mixing as the irreversible diffusive flux across isopycnals that arises due to macroscopic fluid motions, as in \cite{peltier_mixing_2003}.
This flux is sometimes expressed as the mean vertical flux of density, or equivalently buoyancy ${b=-g(\rho-\rho_0)/\rho_0}$.
The flux $\langle w^\prime b^\prime\rangle$ can however include significant contributions from entirely reversible processes such as internal waves.
Indeed equating buoyancy flux and irreversible mixing is only appropriate when both are averaged over time and applied to a statistically stationary state.
\cite{winters_available_1995} show that the true rate of irreversible, diapycnal mixing in a Boussinesq fluid is equal to the conversion rate of available potential energy (APE) to background potential energy (BPE).
As introduced by \cite{lorenz_available_1955}, APE refers to the change in energy resulting from adiabatically sorting the buoyancy field of a fluid to its state of minimum potential energy.
By extending the APE framework to compressible flows \cite{tailleux_energetics_2009} argues that mixing should in fact be described as the dissipation of APE into internal energy, which is balanced exactly by an enhancement in the generation of BPE in the Boussinesq limit.
In this study, we focus on the dynamics of a single-component Boussinesq fluid with a linear equation of state, and refer the reader to \cite{tailleux_energetics_2009,tailleux_available_2013} for a discussion of mixing and APE in more complex scenarios.

Although the \cite{winters_available_1995} framework provides an exact expression with which to calculate diapycnal mixing, it is not practical for use in oceanographic observations.
The most precise observational estimates of mixing come from vertical microstructure profilers that record small-scale gradients of velocity or temperature in isolated vertical profiles.
The methods of \cite{osborn_oceanic_1972} and \cite{osborn_estimates_1980}, which are derived from mean balances in the buoyancy variance and turbulent kinetic energy equations respectively, can then be used to estimate an effective diapycnal diffusivity $K_d$.
This diffusivity is related to the mixing rate through $\mathcal{M} = K_d N^2$ where $N$ is some appropriate measure of the buoyancy frequency.
Note that $N$ may not be straightforward to identify if there is significant spatio-temporal variability in the flow \citep{arthur_how_2017}.
Both estimation methods are derived on the assumption that the flow is statistically steady and thus that the mixing is well described by some average of the buoyancy flux.
The diffusivity $K_d$ obtained from these microstructure measurements can then be checked against estimates of diffusivity from tracer release experiments \citep{ledwell_evidence_2000}.
Understanding how $K_d$ varies throughout the ocean is also vital for improving the accuracy of global circulation models, where diapycnal turbulent fluxes are only captured through a prescribed parameterization of $K_d$, such as that of \cite{klymak_simple_2010}.

Accurately quantifying mixing in computational fluid dynamics requires the use of direct numerical simulations (DNS) that resolve down to the dissipative scales of motion.
These simulations can then be used to test the assumptions used to derive the above models \citep[as in][]{taylor_testing_2019}, or to quantify the differences in inferred diffusivity arising from the models \citep{salehipour_diapycnal_2015}.
The need to resolve the smallest scales of motion restricts the Reynolds numbers $Re$ it is possible to attain through DNS, and so massive computational grids are needed to push $Re$ up towards geophysical values.
Since the earliest days of simulating turbulence through DNS, triply periodic domains have been used to reduce computational cost \citep{orszag_numerical_1972}.
The lack of fixed boundaries in this setup means that higher values of $Re$ can be obtained.  Thin boundary layers do not need to be resolved and highly efficient pseudospectral methods, exploiting the imposed periodicity,  can be implemented.

\cite{riley_direct_1981} were the first to include a mean density stratification in such a triply periodic setup by decomposing the buoyancy field into a linear profile $N_0^2 z$ and a periodic perturbation $\theta$.
This system has since proved popular for studying the dynamics of high $Re$ stratified turbulence \citep[e.g.][]{staquet_statistical_1998,riley_dynamics_2003,brethouwer_scaling_2007}.
Investigations of mixing in periodic domains, recent examples of which can be found in \cite{maffioli_mixing_2016} and \cite{garanaik_inference_2019}, do not however implement the rigorous \cite{winters_available_1995} framework for quantifying APE, thus identifying explicitly irreversible mixing.
It is common instead to describe diapycnal mixing in terms of the destruction rate of buoyancy variance $\chi$.
Indeed, destruction of variance is often how one would quantify the mixing of a passive scalar \citep[e.g.][]{villermaux_mixing_2019}.
The buoyancy variance also acts as a small-amplitude approximation to the APE, and its dissipation rate $\chi$ has long been of use in field measurements \citep{osborn_oceanic_1972,oakey_determination_1982,gargett_dissipation_1984}.

As we later explore in \S\ref{sec:chi_estimate}, approximating mixing with $\chi$ can result in an over/under-estimate depending on whether the most intense turbulence in the flow preferentially samples regions of locally high/low stratification (and thus is associated with different characteristic \emph{local} values of the buoyancy frequency).
It is therefore important to be able to quantify mixing accurately in the periodic system and identify whether such discrepancies can be significant.
Since the buoyancy in the system is only defined through its periodic perturbation $\theta$, ambiguity arises in how to construct the background state of minimum potential energy.
In \S\ref{sec:quantifying_mixing} we use a simple example to highlight this issue and then provide an extension to the framework of \cite{winters_available_1995} that resolves the ambiguity in the case of triply periodic domains.
\S\ref{sec:simulations} gives a brief overview of the numerical simulations we shall use to test the new framework, including the numerical method used.
\S\ref{sec:results} uses the new framework to analyse the simulations, and compares the exact mixing rates to commonly used estimates.
Finally, we conclude and discuss these results in \S\ref{sec:discussion}, with a particular focus on how our findings impact the estimation and parameterization of mixing in the ocean.

\section{Quantifying mixing in triply-periodic domains} \label{sec:quantifying_mixing}

We consider the problem of quantifying irreversible mixing in a system governed by the dimensionless Boussinesq equations subject to an imposed, constant, mean stratification.
We decompose the buoyancy field as $b=z+\theta$, where $b=z$ represents the buoyancy profile of the imposed mean stratification.
Note that $b$ has been non-dimensionalized by $L_0N_0^2$, where $L_0$ is a typical length scale and $N_0$ is the mean dimensional buoyancy frequency, so the mean buoyancy gradient in the dimensionless system is always equal to one.
\begin{align}
    \del \cdot \bu &= 0 , \label{eq:NS1}\\
    \frac{\pd \bu}{\pd t} + \left(\bu\cdot\del\right)\bu &= -\del p + \frac{1}{Re}\nabla^2\bu + Ri_0 \theta\symbfit{\hat{z}} , \label{eq:NS2}\\
    \frac{\pd \theta}{\pd t} + \left(\bu\cdot\del\right)\theta &= \frac{1}{RePr}\nabla^2\theta - w . \label{eq:NS3}
\end{align}
We apply periodic boundary conditions in all three directions to the flow velocity $\bu$, pressure $p$ and buoyancy perturbation $\theta$. The (dimensionless) domain sizes in the $x$, $y$, and $z$ directions are $L_x$, $L_y$, and $L_z$, respectively. The dimensionless parameters in the system are the Reynolds number, Prandtl number and bulk Richardson number, given by
\begin{align}
    Re &= \frac{L_0U_0}{\nu}, &
    Pr &= \frac{\nu}{\kappa}, &
    Ri_0 &= \frac{{N_0}^2 {L_0}^2}{{U_0}^2} , \label{eq:param_def}
\end{align}
where $U_0$ is a velocity scale, $\nu$ is the kinematic viscosity, and $\kappa$ is the diffusivity of buoyancy.
As mentioned in the introduction, these equations are frequently used in studies of stratified turbulence where the periodicity allows for the use of efficient spectral methods and removes the effect of solid boundaries.

Although the buoyancy perturbation $\theta$ is periodic in the vertical, the total buoyancy ${b=z+\theta}$ is not.
We are instead left with a jump condition for $b$ at the upper and lower boundaries that has consequences for the calculation of irreversible mixing and potential energy.
The mean potential energy in the domain is defined as
\begin{equation}
    \mathcal{P}(t) = \left\langle -Ri_0 b z \right\rangle, \label{eq:PE_def}
\end{equation}
where $\langle f \rangle = \frac{1}{V}\int_V f \, dV$ denotes an average over the domain volume $V$.
Substituting $\theta = b - z$ into \eqref{eq:NS3} and multiplying by $-Ri_0z$ provides an evolution equation for the potential energy in the form
\begin{equation}
    \frac{d\mathcal{P}}{dt} = -Ri_0\left\langle wb\right\rangle + \frac{Ri_0}{V}\int_{\pd V} zb\bu\cdot \symbfit{n} \,\mathrm{d}S + \frac{Ri_0}{RePr}\left\langle \frac{\pd b}{\pd z}\right\rangle - \frac{Ri_0}{VRePr}\int_{\pd V} z\del b \cdot \symbfit{n} \,\mathrm{d}S . \label{eq:PE_full_evo}
\end{equation}
Taking the top and bottom boundaries to be at $z=L_z$ and $z=0$ respectively, and applying the periodic boundary conditions simplifies the equation to
\begin{equation}
    \frac{d\mathcal{P}}{dt} = -Ri_0 \langle w\theta \rangle + Ri_0 \left.\overline{w\theta}\right|_{z=0} - \frac{Ri_0}{RePr} \left.\frac{\pd\overline{\theta}}{\pd z}\right|_{z=0} , \label{eq:PE_evo_simple}
\end{equation}
where an overbar denotes a horizontal average, defined as $\overline{f} = \frac{1}{A}\iint_A f \dA$ where $A$ is the cross-sectional area of the domain and $\dA = \mathrm{d}x\mathrm{d}y$ is the area element.
The conversion rate of internal energy to potential energy, given by the third term on the right hand side of \eqref{eq:PE_full_evo}, has been cancelled out by the main contribution of the diffusive flux through the boundary - the final term in \eqref{eq:PE_full_evo}.
The evolution equation \eqref{eq:PE_evo_simple} highlights how sensitive the evolution of the potential energy can be to the choice of the boundary.

The accurate quantification of irreversible mixing requires partitioning the potential energy into background and available components.
The background potential energy (BPE) is defined as the minimum value of potential energy that can be achieved through adiabatic rearrangement of the fluid in the domain.
In this minimum state, the buoyancy profile is given by a monotonically increasing one-dimensional function $b_*(z,t)$, so the mean BPE is given by $\mathcal{P}_B = \langle-Ri_0b_*z\rangle$.
\cite{winters_available_1995} show that BPE can also be expressed as
\begin{equation}
    \mathcal{P}_B(t) = \left\langle -Ri_0 \, b(\bx, t) \, z_*(\bx,t) \right\rangle, \label{eq:Winters_BPE}
\end{equation}
where $z_*$ is the height a parcel of fluid with buoyancy $b(\bx,t)$ is moved to under the adiabatic rearrangement. Following \cite{lorenz_available_1955}, the available potential energy (APE) is defined as the surplus potential energy
\begin{equation}
    \mathcal{P}_A(t) = \left\langle -Ri_0 b \left(z - z_*\right)\right\rangle . \label{eq:Winters_APE}
\end{equation}
The rate of irreversible mixing associated with fluid motion is then given by the energy transfer rate from APE to BPE, which takes the form
\begin{equation}
    \mathcal{M} = \frac{Ri_0}{RePr}\left\langle \left.\frac{\pd Z_*}{\pd b}\right|_{b(\bx,t)} \left|\del b\right|^2 - \frac{\pd b}{\pd z}\right\rangle = \frac{Ri_0}{RePr}\left(\left\langle \left.\frac{\pd Z_*}{\pd b}\right|_{b(\bx,t)} \left|\del b\right|^2\right\rangle - 1\right), \label{eq:Mix_def}
\end{equation}
where $Z_*(b,t)$ is the inverse function associated with the sorted buoyancy profile $b_*$ which satisfies $z_*(\bx, t) = Z_*(b(\bx,t), t)$.
It is important to appreciate that the term scaling $|\del b|^2$ in \eqref{eq:Mix_def} is effectively the inverse square of the buoyancy frequency of the sorted variables, and so accentuates the contributions where the sorted buoyancy gradient is relatively weak. 
As discussed below in \S\ref{sec:chi_estimate}, this is a potential source of difference between $\mathcal{M}$ and the buoyancy variance destruction rate $\chi$.

Note that \citet{tailleux_available_2013} argues for a more general definition of APE, where $b_*$ is replaced by an arbitrary reference state $b_r$ that can depend on a wide range of thermodynamic quantities.
This definition is particularly useful for its possible extension to multicomponent, compressible fluids as shown by \citet{tailleux_local_2018}.
Defining buoyancy relative to an arbitrary reference state also highlights an inherent ambiguity in calculating APE.
If APE is quantified as the cumulative work done against buoyancy forces, then different definitions of a reference buoyancy state may alter how one interprets mixing and APE dissipation.
For the single-component, Boussinesq, linear equation of state considered here, we shall continue to use the adiabatically sorted $b_*$ given its aforementioned connection to a widely-studied approach used for the quantification of diapycnal mixing.

\begin{figure}
    \centering
    \includegraphics[width=\textwidth]{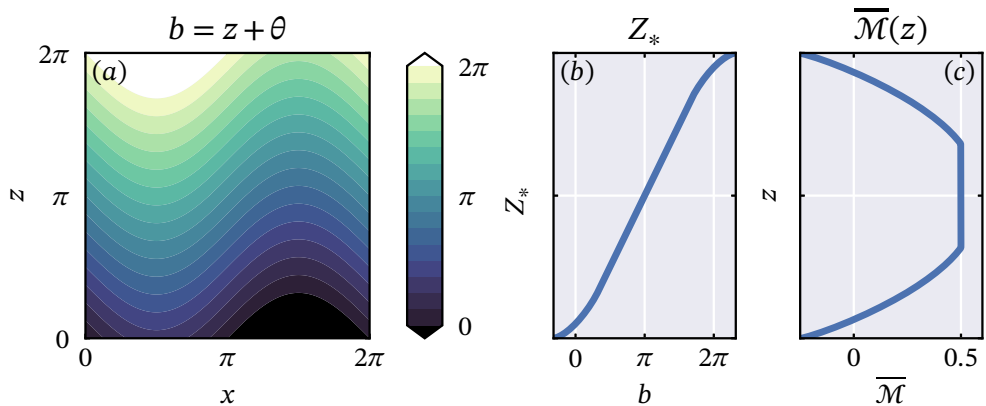}
    \caption{
        $(a)$ displays contours of the total buoyancy field given by $\theta=\sin x$;
        $(b)$ shows the sorted profile $Z_*(b)$ associated with this buoyancy field;
        $(c)$ shows the horizontally averaged irreversible mixing rate $\overline{\mathcal{M}}(z) = \overline{\frac{\pd Z_\ast}{\pd b} |\del b|^2} - \frac{\pd \overline{b}}{\pd z}$.
        Note that an overbar here denotes an average over $x$, and $\pd Z_\ast/\pd b$ is evaluated at $b(x,z)$.
    }
    \label{fig:aperiodic_example}
\end{figure}

We now present a simple example to highlight how the aperiodicity of $b$ can cause issues for calculating the mixing rate $\mathcal{M}$.
We consider the buoyancy field given by $\theta = \sin x$ in a domain of length $2\pi$.
This might be thought of as a representation of the buoyancy field associated with a standing internal gravity wave, at an instant when half the columns of fluid in the domain are raised up and half are pushed down relative to their equilibrium location.

The total buoyancy field $b=z+\sin x$ and its corresponding sorted profile $Z_*(b)$ are shown in figures \ref{fig:aperiodic_example}a and \ref{fig:aperiodic_example}b respectively.
In an unbounded domain, we would expect a linear profile for $Z_*$ since the wave is simply a rearrangement of the initial linear stratification.
However by taking the boundaries at $z=0$ and $z=2\pi$, we produce a profile with deviations from a uniform slope close to these values.
Since $\theta$ is independent of $z$, we would also expect the mixing rate $\mathcal{M}$ to be constant regardless of the vertical extent that we average over.
Figure \ref{fig:aperiodic_example}c instead shows that the variations in $\pd Z_*/\pd b$ change the value of $\mathcal{M}$ across much of the domain, with the horizontally-averaged mixing rate even taking negative values close to the boundary.

This issue has caused problems in the literature before.
By sorting the buoyancy profile of a rectangular, periodic domain, \citet{bouruet-aubertot_particle_2001} observe extremely large oscillations in the APE of a breaking internal wave (as shown in figure 9a of that paper) due to fluxes across the top/bottom boundary of the domain.
The large oscillations make it difficult to draw precise, quantitative conclusions about the evolution of APE and diapycnal mixing in that system.

\subsection{Potential energy between isopycnal boundaries \label{sec:isopycnal_PE}}

\begin{figure}
    \centering
    \includegraphics[width=\linewidth]{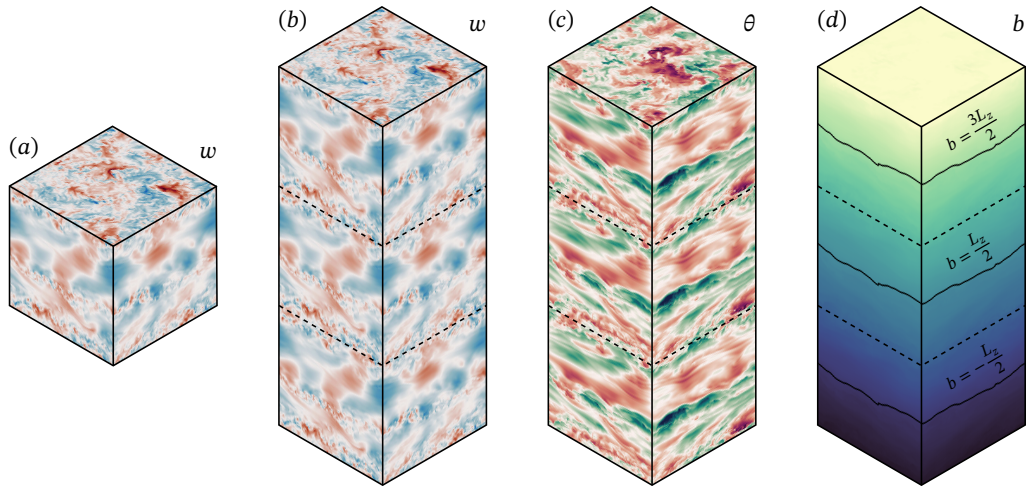}
    \caption{
        Example schematic of tiling the periodic computational domain vertically.
        The vertical velocity $w$, shown in $(a)$ and $(b)$, and buoyancy perturbation $\theta$, shown in $(c)$, simply repeat thanks to their periodic boundary conditions.
        The total buoyancy $b=z+\theta$, shown in $(d)$, is not periodic in the vertical, although isopycnal surfaces separated by the vertical period $L_z$ are of identical shape.
    }
    \label{fig:stacking}
\end{figure}

We propose the use of a control volume bounded by surfaces of constant buoyancy (isopycnals) to tackle the highlighted issue of quantifying mixing in triply-periodic domains.
Consider tiling the computational domain by stacking several computational domains vertically, as in figure \ref{fig:stacking}.
The velocity and buoyancy perturbation repeat in each domain, but the vertical coordinate, $z$, is continuous such that the total buoyancy in one tile is $L_z$ larger than the total buoyancy at the same relative position in the tile immediately below it.
In this system it is particularly useful to consider two isopycnals separated by the vertical periodic length, i.e. $L_z$.
These isopycnals will have the same shape due to the periodicity of $\theta$, and the volume enclosed by these two isopycnals will therefore be constant.
The buoyancy profile can then be sorted into a \emph{background} state $b_*(z)$, where the parcels are sorted into the one-dimensional domain $0\leq z < L_z$.
Although this background profile must have a mean vertical gradient equal to the imposed mean stratification, its local gradients $\pd b_\ast/\pd z$ can vary more generally.
In the simple example considered in figure \ref{fig:aperiodic_example}, this technique recovers the linear profile $Z_\ast(b)=b$ expected from the column displacement argument mentioned above.

We now describe more precisely the details of implementing isopycnal boundaries for quantifying available potential energy and mixing.
We first choose a buoyancy value $b_0$ that defines the lower boundary surface $z_1(x,y,t)$ implicitly through
\begin{equation}
    b(x, y, z_1(x,y,t), t) = b_0 . \label{eq:z1_def}
\end{equation}
Vertical periodicity of $\theta$ then requires that the upper boundary surface $b=b_0 + L_z$ is defined by ${z_2 = L_z + z_1}$.
It is important to appreciate that \eqref{eq:z1_def} defines $z_1$ (and hence also $z_2$) as a single surface that spans the horizontal cross-section of the domain.
This ensures that the volume enclosed by the isopycnals is clearly defined.
To aid the calculation of volume integrals, we also require (essentially for clarity of exposition) $z_1$ to be a single-valued function of $x$ and $y$, or equivalently that the boundary isopycnal cannot exhibit overturning.
Such an isopycnal may be difficult to find in homogeneous turbulence, although stratified flows are often strongly spatially inhomogeneous.
A discussion of how this approach could be generalised for an overturning isopycnal surface can be found in appendix \ref{app:iso}.

Constructing evolution equations for mean energy quantities involves taking time derivatives of volume integrals.
Since the boundaries of our domain are now time-dependent, we must apply the Leibniz rule to any such integral, that is
\begin{equation}
    \frac{d}{dt}\left( \int_V f \dV \right) = \int_V \frac{\pd f}{\pd t} \dV + \int_A \left( \left. f\right|_{z=z_2} - \left. f\right|_{z=z_1}\right) \frac{\pd z_1}{\pd t} \dA , \label{eq:Leibniz}
\end{equation}
where $A$ is the horizontal cross-sectional area of the domain and the area element $\mathrm{d}A = \mathrm{d}x\mathrm{dy}$.

The mean kinetic energy of the system $\mathcal{K} = \langle |\bu|^2\rangle /2$ is unaffected by the change of boundaries, since its integrand is periodic in the vertical direction.
The evolution of $\mathcal{K}$ can therefore be derived straightforwardly from \eqref{eq:NS2} by applying \eqref{eq:NS1} to obtain the simple form
\begin{equation}
    \frac{d\mathcal{K}}{dt} = \J - \eps , \label{eq:K_evo}
\end{equation}
where the buoyancy flux and kinetic energy dissipation rate are respectively given by
\begin{align}
    \J &= Ri_0 \langle w\theta \rangle, &
    \eps &= \frac{1}{Re} \left\langle \frac{\pd u_i}{\pd x_j} \frac{\pd u_i}{\pd x_j} \right\rangle. \label{eq:J_eps_def}
\end{align}
Note that from this definition, positive values of buoyancy flux correspond to a conversion of potential energy to kinetic energy.

However, extra terms do arise compared to \eqref{eq:PE_full_evo} when deriving the potential energy evolution equation.
These new terms provide a secondary reservoir of potential energy for the system, as is explained below.
The advective flux across the boundary, given by the second term on the right of \eqref{eq:PE_full_evo}, is now zero since the bounding isopycnals have the same shape and the same gradients due to periodicity.
Applying the Leibniz result \eqref{eq:Leibniz} to the potential energy $\mathcal{P}$ and imposing the boundary conditions therefore produces the evolution equation
\begin{equation}
    \frac{d\P}{dt} + \frac{d\mathcal{S}}{dt} = - \J + \mathcal{D}_p - \mathcal{F}_d.  \label{eq:P+S_evo}
\end{equation}
(A more detailed derivation of this equation can be found in appendix \ref{app:tot_PE}.)
$\mathcal{D}_p=Ri_0/RePr$ is the conversion rate of internal energy to potential energy, and $\mathcal{F}_d$ is the diffusive boundary term given by
\begin{equation}
    \mathcal{F}_d = \frac{Ri_0}{ARePr} \int_A \left[ \frac{|\del b|^2}{\pd b/\pd z}\right]_{z=z_1} \dA = \frac{Ri_0}{RePr} \overline{\left( \frac{|\del b|^2}{\pd b/\pd z}\right)_{z=z_1}} , \label{eq:flux_def}
\end{equation}
where the overbar denotes a cross-sectional average over $x$ and $y$, importantly taken after the quantity in brackets is evaluated at $z=z_1(x,y,t)$.
We refer to the quantity $\mathcal{S}$ as the \emph{surface potential energy}, where $\mathcal{S}$ is  defined as
\begin{equation}
    \mathcal{S} = Ri_0\left( \langle b_0 z \rangle + \frac{\overline{{z_2}^2}}{2}\right) . \label{eq:S_def}
\end{equation}
We can arbitrarily set $b_0=0$ in all of the above by shifting our vertical coordinate to $z-b_0$.
$\mathcal{S}$ then takes the form of potential energy associated with an interface at $z_2$, motivating our choice for its name.

\subsection{APE and BPE between isopycnal boundaries \label{sec:isopycnal_APE}}

\begin{figure}
    \centering
    \includegraphics[width=\textwidth]{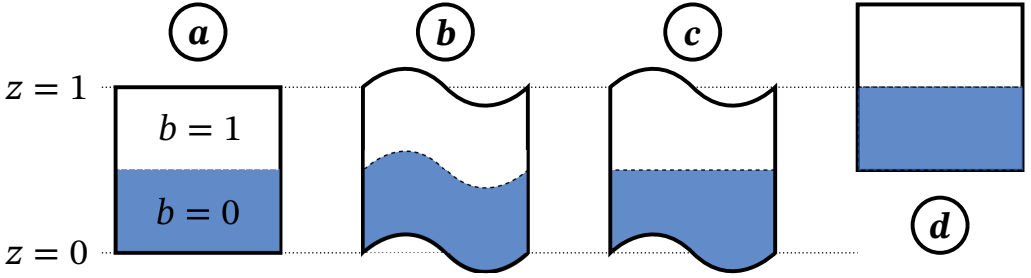}
    \caption{A sketch of two-layer buoyancy fields with varying vertical boundaries.}
    \label{fig:APE_sketch}
\end{figure}

Using the \cite{winters_available_1995} form of APE defined in \eqref{eq:Winters_APE} is not appropriate for the time-varying domains considered here.
This can be understood by considering the simple two-layer system shown in figure \ref{fig:APE_sketch}.
Panel $(a)$ shows the background state obtained through constructing the one-dimensional buoyancy profile $b_*$ for the buoyancy fields in panels $(b)$ and $(c)$.
Since the buoyancy field in figure \ref{fig:APE_sketch}b can be obtained from the background state through shifting the same number of fluid columns up as down, $\mathcal{P}$ does not change between states $(a)$ and $(b)$.
$\P = \P_B$ therefore holds for state $(b)$, and hence $\P_A=0$.
It is simple however to construct a state $(c)$ with lower potential energy than state $(b)$.
The \cite{winters_available_1995} definition would then in fact give $\P_A < 0$ for the buoyancy profile in figure \ref{fig:APE_sketch}c, which is not consistent with the concept of available potential energy.

We aim to define a new APE variable $\mathcal{A}$ that can be used in the time-varying domain.
Progress can be made by considering the \emph{total potential energy} $\mathcal{P}+\mathcal{S}$ that appears in \eqref{eq:P+S_evo}.
The decrease in $\P$ from figure \ref{fig:APE_sketch}a to \ref{fig:APE_sketch}c is matched exactly by an increase in $\mathcal{S}$.
In terms of the total potential energy, states $(a)$ and $(c)$ are therefore equivalent background states.
This motivates subdividing the potential energy into
\begin{equation}
    \P + \mathcal{S} = \mathcal{A} + \mathcal{B} . \label{eq:P+S_split}
\end{equation}
We expect $\mathcal{A}=0$ for states $(a)$, $(c)$, and $(d)$ in figure \ref{fig:APE_sketch}.
In particular for state $(d)$ this means that any change in $\P+\mathcal{S}$ due to a vertical shift of the domain is captured by the background potential energy $\mathcal{B}$.
We therefore construct the background profile $b_*(z)$ over the domain $\overline{z_1}<z<\overline{z_2}$, such that
\begin{align}
    Z_*(0,t) &= \overline{z_1}(t), & Z_*(L_z,t) &= \overline{z_2}(t), & b_*(\overline{z_1},t) &= 0, & b_*(\overline{z_2},t) &= L_z.
    \label{eq:Zstar_BCs}
\end{align}
This ensures that any change in $\mathcal{P}$ due to a shift in the mean height of the lower isopycnal $\overline{z_1}$ leads to a corresponding change in $\mathcal{P}_B$.
Accounting also for the corresponding change in $\mathcal{S}$ leads to the following definitions for background and available potential energy:
\begin{align}
    \mathcal{B} &= \left\langle -Ri_0 b z_\ast \right\rangle + \frac{Ri_0}{2} \overline{z_2}^{\,2} , \label{eq:B_def}\\
    \mathcal{A} &= \left\langle -Ri_0 b \left(z-z_\ast\right) \right\rangle + \frac{Ri_0}{2} \left( \overline{{z_2}^2} - \overline{z_2}^{\,2}\right) . \label{eq:A_def}
\end{align}
Note that for a closed system with fixed, insulated boundaries, these definitions recover the \cite{winters_available_1995} form for BPE and APE up to a constant in the BPE.

Evolution equations for these quantities can be readily obtained through multiplying the buoyancy evolution equation \eqref{eq:NS3} by $z_*$ and taking volume averages.
An analogous derivation as that leading to \eqref{eq:P+S_evo}, as shown in appendix \ref{app:BPE}, results in
\begin{equation}
    \frac{d\mathcal{B}}{dt} = \mathcal{M} + \mathcal{D}_p - \mathcal{F}_d, \label{eq:B_evo}
\end{equation}
where $\mathcal{M}$ is the irreversible mixing rate defined in \eqref{eq:Mix_def}.
Subtracting \eqref{eq:B_evo} from \eqref{eq:P+S_evo} also gives an evolution equation for our new APE variable as
\begin{equation}
    \frac{d\mathcal{A}}{dt} = -\J - \mathcal{M} . \label{eq:A_evo}
\end{equation}
We therefore recover the simple evolution equation for APE in a closed system, where the irreversible mixing rate $\mathcal{M}$ may also be identified with a destruction of APE \citep[e.g.][]{peltier_mixing_2003}.

\subsection{Comparison to local APE of Scotti \& White (2014) \label{sec:localAPE}}

The concept of local APE is used as an alternative framework for quantifying available potential energy in situations where fluxes through a boundary are important.
Originally devised by \citet{holliday_potential_1981} and \citet{andrews_note_1981}, local APE quantifies the work done against buoyancy forces to move a fluid parcel from a reference position to its actual position.
The framework has seen renewed interest recently in its application to numerical simulations.
We follow \cite{scotti_diagnosing_2014} in defining the local APE density $E_{APE}$ as a function of space and time by
\begin{equation}
    E_{APE}(\bx,t) = -Ri_0 \int_{b_*(z,t)}^{b(\bx,t)} z - Z_*(s,t) \,\mathrm{d}s . \label{eq:locAPE_def}
\end{equation}
We use this form primarily for its ease of notation, although as we show in appendix \ref{app:APE_equivalence}, for the setup we consider it is equivalent to various other expressions proposed for local APE density.
Although this quantity varies in space and time, its dependence on the globally sorted profiles $b_*$ and $Z_*$ means that it cannot be calculated solely from local fields.
In particular, the issue for quantifying mixing highlighted by figure \ref{fig:aperiodic_example} remains unless we change the domain over which $b_*$ is constructed.
If we instead take isopycnal boundaries as in \S\ref{sec:isopycnal_PE}, then $E_{APE}$ becomes a useful tool for investigating the local mechanisms that lead to mixing in the domain.

Indeed, we can relate the volume-averaged $E_{APE}$ to our global APE variable $\mathcal{A}$ as follows.
The mean local APE defined in this way can also be written in the form
\begin{equation}
    E_A \equiv \left\langle E_{APE} \right\rangle = -Ri_0 \left\langle b \left( z-z_\ast \right) \right\rangle - Ri_0 \left\langle \int_z^{z_\ast(\bx,t)} b_\ast (s,t) \,\mathrm{d}s \right\rangle . \label{eq:mean_locAPE}
\end{equation}
\cite{winters_available_2013} explain that the final term in this expression accounts for the energy changes arising from the requirement of incompressibility, leading to displacement of some fluid elements to make room for the rearrangement of a fluid parcel in the sorting process.
They also showed through considering fluid parcel exchanges that this term vanishes in the case of fixed horizontal boundaries.

We now consider a simple example to show how this term can change with non-horizontal boundaries and how it relates to the additional terms in \eqref{eq:A_def}.
We take $\theta = - z_1(x,y,t)$ as the buoyancy perturbation field, so the domain represents that of a uniform stratification where each fluid column has been shifted so that the $b=0$ isopycnal is at $z_1$, analogously to the situations shown in figures \ref{fig:aperiodic_example}a and \ref{fig:APE_sketch}b.
In this case, the reference profiles simply take the form $b_*(s,t)=s-\overline{z_1}(t)$, and $Z_*(s,t) = s+\overline{z_1}(t)$. 
We can therefore analytically compute
\begin{equation}
    E_{APE}(\bx,t) = \frac{Ri_0}{2} \left(z_1(x,y,t) - \overline{z_1}(t)\right)^2 .
\end{equation}
Considering each of the terms in \eqref{eq:mean_locAPE} separately, we also find that
\begin{align}
    -Ri_0\left\langle b\left(z-z_\ast\right) \right\rangle &= 0, &
    -Ri_0\left\langle \int_z^{z_\ast(\bx,t)} b_\ast (s,t) \,\mathrm{d}s \right\rangle &= \frac{Ri_0}{2}\left(\overline{{z_1}^2} - \overline{z_1}^{\,2}\right). 
\end{align}
The integral term therefore accounts for all of the available potential energy associated with the surface potential energy $\mathcal{S}$.
Since $\overline{{z_2}^2} - \overline{z_2}^{\,2} = \overline{{z_1}^2} - \overline{z_1}^{\,2}$, we find that the volume-integrated local APE exactly matches the changes we propose for the global APE in \eqref{eq:A_def}.

In contrast to $\mathcal{A}$, the calculation of $E_{APE}$ does not rely on computing a surface integral over the isopycnal boundary.
Indeed the moving boundary only affects the calculation of local APE through the boundary conditions \eqref{eq:Zstar_BCs} for the reference profiles $b_*(z,t)$ and $Z_*(s,t)$, and even here one only needs to know the mean height of the isopycnal.
The strong agreement between $\mathcal{A}$ and $E_A$ gives us hope that in flows where $\mathcal{A}$ is not well defined, $E_A$ can provide an accurate measure of available potential energy.

\section{Numerical simulations} \label{sec:simulations}

\begin{table}
    \centering
    \label{tab:simulations}
    \begin{tabular}{l *{7}{c}}
        \toprule
        Simulation & F1 & F2 & F3 & U1 & U2 & U3 & U4 \\
        \midrule
        Reynolds number ($Re$) & $10\,000$ & $10\,000$ & $10\,000$ & $8000$ & $8000$ & $5000$ & $5000$ \\
        Domain size ($L_x \times L_y \times L_z$) & \multicolumn{3}{c}{$2\pi \times 2\pi \times 2\pi$} & \multicolumn{4}{c}{$8\pi \times \pi/2 \times 2\pi$} \\
        Resolution & \multicolumn{3}{c}{$1024 \times 1024 \times 1024$} &\multicolumn{4}{c}{$4096 \times 256 \times 1024$} \\
        Initial condition & \multicolumn{3}{c}{IGW spectrum} & \multicolumn{4}{c}{Shear and wave} \\
        Forcing & Vortical & Waves & Waves & \multicolumn{4}{c}{Unforced} \\
        \bottomrule
    \end{tabular}
    \caption{Overview of the various numerical simulations, all of which are performed using a bulk Richardson number $Ri_0=1$ and Prandtl number $Pr=1$.}
\end{table}

We apply the extended APE framework developed in \S\ref{sec:isopycnal_APE} to two sets of direct numerical simulations.
All of these simulations are performed using \textsc{Diablo}, which uses a third-order Runge--Kutta scheme for time stepping and a pseudo-spectral method for calculating spatial derivatives \citep{taylor_numerical_2008}.
The software also implements dealiasing of nonlinear terms through a $2/3$ rule.
One set of simulations (set F) adds forcing terms to \eqref{eq:NS2} and \eqref{eq:NS3} to produce a flow in a statistically steady state, whereas the other simulations (in set U) solve the equations unforced as an initial value problem.
All of the simulations considered here are performed using a bulk Richardson number $Ri_0=1$ and Prandtl number $Pr=1$.

The first set of simulations are those used in our previous study on mixing in forced stratified turbulence \citep{howland_mixing_2020}.
We refer to the simulations H, R and P in that paper by F1, F2, and F3 respectively, and outline some of their key parameters in table \ref{tab:simulations}.
Simulation F1 is forced by randomly phased large-scale vortical modes, and importantly features no direct forcing of the buoyancy field.
The evolution equations \eqref{eq:B_evo} and \eqref{eq:A_evo} for $\mathcal{B}$ and $\mathcal{A}$ still therefore hold.
On the other hand, simulations F2 and F3 are forced by large-scale internal gravity waves that include a buoyancy forcing component.
The buoyancy forcing can act as a source or sink of potential energy, modifying the evolution equations.
However if we are primarily concerned with diapycnal mixing, it remains useful to calculate the irreversible mixing rate $\mathcal{M}$ in these cases.
For more precise details of the forcing in these simulations, we refer the reader to \cite{howland_mixing_2020}.

\begin{figure}
    \centering
    \includegraphics[width=\textwidth]{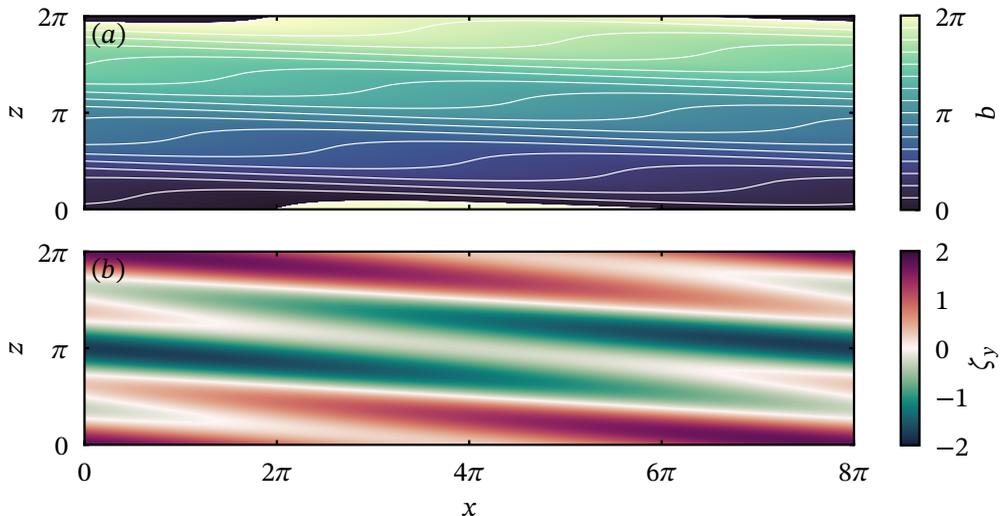}
    \caption{
        The initial condition of simulation U4, where $s=0.75$.
        $(a)$ Contours and colour map of the total buoyancy field $b=z+\theta \ \mathrm{mod} \ 2\pi$.
        $(b)$ Colour map of the spanwise vorticity $\zeta_y = \frac{\pd u}{\pd z} - \frac{\pd w}{\pd x}$.
        }
    \label{fig:IC_s075}
\end{figure}

The second set of simulations investigate the interaction of a sinusoidal vertical shear flow and a plane internal gravity wave. 
The initial velocity and buoyancy fields are given by ${\bu = \left(\sin z \right)\widehat{\symbfit{x}} + \bu^\prime}$ and $\theta = \theta^\prime$ respectively, where
\begin{align}
    \theta^\prime &= \frac{s}{m} \cos\left(kx+mz\right), &
    \bu^\prime &= \frac{s}{\sqrt{k^2+m^2}} \sin\left(kx+mz\right) \left( 1, 0, -\frac{k}{m}\right).
    \label{eq:wave_IC}
\end{align}
We express the initial amplitude of the internal wave through its steepness $s$ and choose the wave vector $\symbfit{k}=(k,l,m)=(1/4, 0, 3)$ based on the typical aspect ratios of waves observed in the thermocline by \cite{alford_observations_2000}.
Small-amplitude noise is added to the initial velocity field to allow the development of three-dimensional motion from the two-dimensional initial condition.
Simulations U1 and U3 use an initial wave steepness of $s=1$, with $s=0.5$ for simulation U2 and $s=0.75$ for simulation U4.
As an example, the initial buoyancy and spanwise vorticity fields for U4 are shown in figure \ref{fig:IC_s075}.
Note that by taking ${b=z+\theta \ \mathrm{mod} \ 2\pi}$ in figure \ref{fig:IC_s075}a, we have effectively defined isopycnal boundaries at $b=0$ and $b=2\pi$.
(A more detailed  analysis of the properties of these simulations is presented in \cite{howland_shear-induced_2021}.)

\section{Results} \label{sec:results}
\subsection{Energy budgets} \label{sec:energy_budgets}

\begin{figure}
    \centering
    \includegraphics[width=\textwidth]{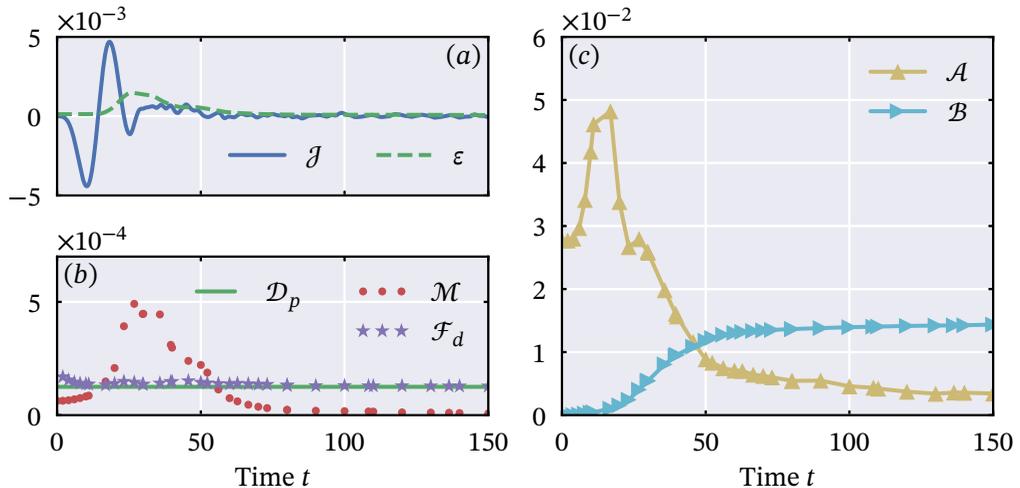}
    \caption{
        Energy budgets for simulation U1.
        $(a)$ Time series of the mean buoyancy flux and viscous dissipation rate;
        $(b)$ time series of the BPE budget terms;
        $(c)$ time series of available and background potential energies defined in \eqref{eq:B_def} and \eqref{eq:A_def}.
        The time series for $\mathcal{B}$ is shifted by $-\mathcal{B}(0)$ for clarity.
        Terms denoted by symbols are computed from full flow output files, and so have lower time resolution than $\J$ and $\eps$, which are computed `on-the-fly'.
    }
    \label{fig:PE_budget_U1}
\end{figure}

We now investigate the evolution of background and available potential energy in the various simulations, and consider how terms in the energy budgets \eqref{eq:B_evo} and \eqref{eq:A_evo} relate to the flow dynamics.
Figure \ref{fig:PE_budget_U1} plots a range of time series associated with the unforced simulation U1.
The kinetic energy budget terms $\J$ and $\eps$, defined in \eqref{eq:J_eps_def}, are shown in figure \ref{fig:PE_budget_U1}a, and the BPE budget terms from \eqref{eq:B_evo} are shown in figure \ref{fig:PE_budget_U1}b.
Time series of $\mathcal{A}$ and $\mathcal{B}$ are finally shown in figure \ref{fig:PE_budget_U1}c.
Up to time $t\approx20$, the energetics are dominated by large, reversible changes through the buoyancy flux.
The initial increase in $\mathcal{A}$ seen in figure \ref{fig:PE_budget_U1}c is almost entirely returned to the kinetic energy through wave-mean flow interactions.
During this time, there is little mixing and any changes in $\mathcal{B}$ are small.
A wave breaking event follows, producing an intermittent burst of turbulent activity that coincides with high values of the diapycnal mixing rate $\mathcal{M}$ and the KE dissipation rate $\eps$.
For $30<t<50$, this mixing coincides with positive values of the mean buoyancy flux, leading to a fast drop in $\mathcal{A}$.
The flow relaminarizes at late times, with all quantities tending to constant values and small fluctuations persisting in the APE and buoyancy flux.

\begin{figure}
    \centering
    \includegraphics[width=\textwidth]{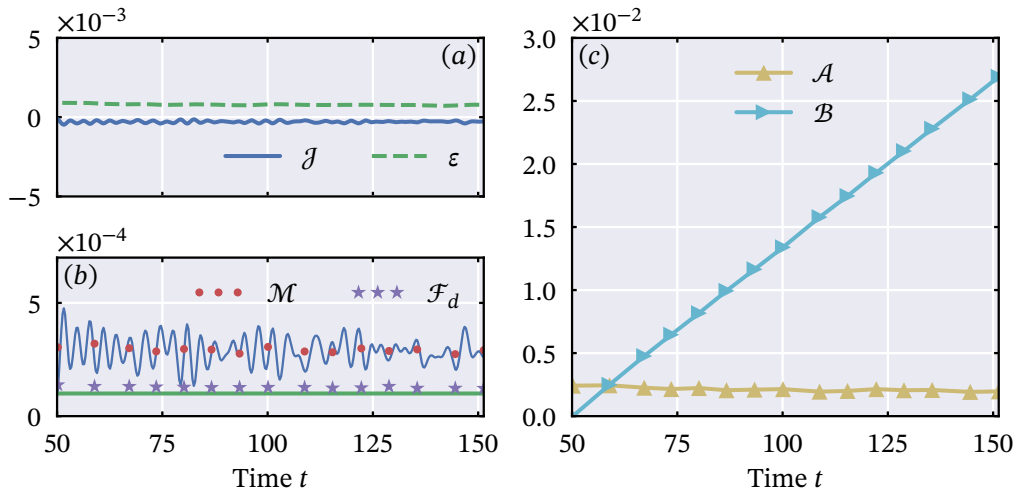}
    \caption{
        Potential energy budgets for the late-time statistically steady state achieved in simulation F1.
        Subplots are as in figure \ref{fig:PE_budget_U1}, with $-\J$ additionally plotted on panel $(b)$.
    }
    \label{fig:PE_budget_F1}
\end{figure}

The increase in $\mathcal{B}$ over the full duration of the simulation is well described by the total diapycnal mixing associated with the breaking event.
Indeed the other non-negligible terms in the budget \eqref{eq:B_evo} are close to being equal, as shown in both figures \ref{fig:PE_budget_U1} and \ref{fig:PE_budget_F1}.
The diffusive boundary term $\mathcal{F}_d$ primarily acts to cancel out any increase in $\mathcal{B}$ due to the conversion of internal energy to potential energy through $\mathcal{D}_p$.
This cancellation is exact when the boundary has no lateral variation, and arises since the system is forced to maintain a constant mean buoyancy gradient through the periodicity of $\theta$.

Figure \ref{fig:PE_budget_F1} repeats the analysis of figure \ref{fig:PE_budget_U1} for the forced simulation F1.
We only consider the statistically steady period achieved at late times in this flow.
Unlike in the unforced simulation, the mean buoyancy flux remains negative throughout as shown in figure \ref{fig:PE_budget_F1}a, providing a source of APE from the kinetic energy.
Figure \ref{fig:PE_budget_F1}b furthermore shows that the buoyancy flux is on average in balance with the mixing rate, leading to an approximately constant value of $\mathcal{A}$, as shown in figure \ref{fig:PE_budget_F1}c.
The constant mixing rate also predictably leads to a linear increase in the background potential energy.

\subsection{Visualising mixing with local APE} \label{sec:locAPE_results}

\begin{figure}
    \centering
    \includegraphics[width=\textwidth]{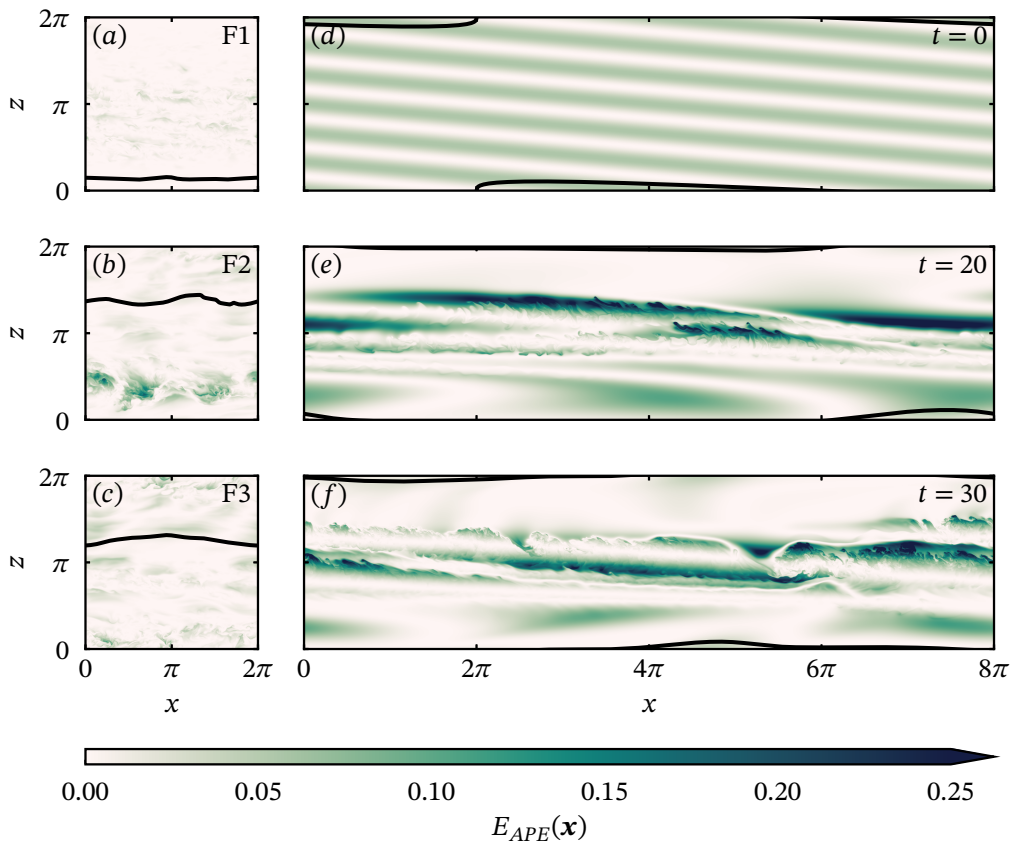}
    \caption{
        Vertical plane snapshots of $E_{APE}$ as defined in \eqref{eq:locAPE_def}.
        Solid lines in each case denote the isopycnal boundary $z_1$ from which the APE is calculated.
        Snapshots from the forced simulations are each taken at time $t\approx 150$ with runs F1, F2, F3 shown in $(a)$, $(b)$, $(c)$ respectively.
        Panels $(d)$-$(f)$ display the evolution of $E_{APE}$ in simulation U1 from the initial condition to the peak in mixing at time $t=30$.
    }
    \label{fig:localAPE}
\end{figure}

We can further investigate the local processes that lead to the global results above by analysing the distribution of local APE throughout the domain.
Figure \ref{fig:localAPE} plots snapshots of $E_{APE}(\bx,t)$ from simulations F1-F3 and from simulation U1 at various times.
Since the turbulence arising in each simulation is patchy and inhomogeneous, we are able to choose appropriate isopycnal boundaries for each simulation and hence calculate the surface potential energy $\mathcal{S}$.
These isopycnal boundaries are shown in figure \ref{fig:localAPE} as solid black lines.

Data from the forced simulations of set F are presented in figures \ref{fig:localAPE}a-c.
Each snapshot of $E_{APE}$ is taken at time $t\approx 150$, when the turbulence is in a statistically steady state.
Figure \ref{fig:localAPE}a highlights low local APE values throughout the domain of simulation F1.
Increased $E_{APE}$ occurs only at small scales and in regions with high turbulent dissipation rates (not shown).
In this sense, APE is primarily associated here with the distortion of the buoyancy field by turbulence, and not with internal waves.
By contrast, figures \ref{fig:localAPE}b and \ref{fig:localAPE}c show patches of high local APE throughout the domain at a range of scales.
This is consistent with associating mixing with intermittent, large-scale overturns and convectively-driven turbulence, as discussed in \citet{howland_mixing_2020}.

The development of local APE during the unforced simulation U1 is presented in figures \ref{fig:localAPE}d-f.
The distribution of $E_{APE}$ in the initial condition is shown in figure \ref{fig:localAPE}d, and is entirely associated with the internal gravity wave described by \eqref{eq:wave_IC}.
At early times, the wave is refracted by the shear flow, leading to a distortion of the banded structure in the local APE field.
By time $t=20$, $E_{APE}$ preferentially accumulates in the upper half of the domain while maintaining some signal of the wave structure, as shown in figure \ref{fig:localAPE}e.
The large values of $E_{APE}$ lead to locally unstable buoyancy profiles, and the development of convective instabilities.
The associated convection converts APE to kinetic energy through the buoyancy flux, and also promotes the emergence of small scale structures seen in figure \ref{fig:localAPE}e.
Later, at $t=30$, the flow becomes more complex with the development of shear-driven turbulent billow structures.
These structures, seen prominently on the right of figure \ref{fig:localAPE}f, span regions of both high and low $E_{APE}$.
Although the volume-averaged mixing rate peaks near this time, the banded structure of $E_{APE}$ leads to strong local variation in local mixing rates within the turbulent patches.
Mixing is high where turbulence and APE coexist, and it cannot occur where there is no APE to remove.

\subsection{Estimating mixing with \texorpdfstring{$\chi$}{χ}} \label{sec:chi_estimate}

In the limit of small buoyancy perturbations from the uniform, imposed buoyancy gradient, available potential energy can be approximated by
\begin{equation}
    \mathcal{\tilde{A}} = \frac{Ri_0}{2}\left\langle \theta^2 \right\rangle . \label{eq:Lorenz_def}
\end{equation}
This quantity satisfies the simple evolution equation
\begin{equation}
    \frac{d\tilde{\mathcal{A}}}{dt} = -\J - \chi , \label{eq:Lorenz_evo}
\end{equation}
where $\chi$ is the rate of destruction of buoyancy variance, i.e. the dissipation rate  defined by
\begin{equation}
    \chi = \frac{Ri_0}{RePr}\left\langle \left|\del \theta\right|^2 \right\rangle = \frac{Ri_0}{RePr} \left\langle \left|\del b\right|^2 - 1 \right\rangle . \label{eq:chi_def}
\end{equation}
Comparing the definitions \eqref{eq:Mix_def} and \eqref{eq:chi_def}, we see that $\chi$ precisely takes the form of the irreversible, diapycnal mixing rate $\mathcal{M}$ only for the specific case where the sorted background buoyancy profile $b_*$ exactly matches the imposed uniform stratification.
Recall that this imposed constant stratification has a dimensionless buoyancy gradient equal to one by construction.
In our simulations, local deviations in the buoyancy field are not always small and we should treat the above approximation with caution.
For example, during the convective phase ($20<t<30$) of simulation U1 there are sizeable regions of the domain with statically unstable buoyancy gradients.
The peak mixing in this case occurs where the (horizontal) mean buoyancy is in a layered state, with `layers' of relatively low stratification separated by `interfaces' of relatively high stratification compared to the imposed constant buoyancy gradient.
Such layered states are observed to arise naturally in turbulent stratified flows, for a wide variety of dynamical reasons \citep[see][for a review]{caulfield_layering_2021}.

\begin{figure}
    \centering
    \includegraphics[width=\textwidth]{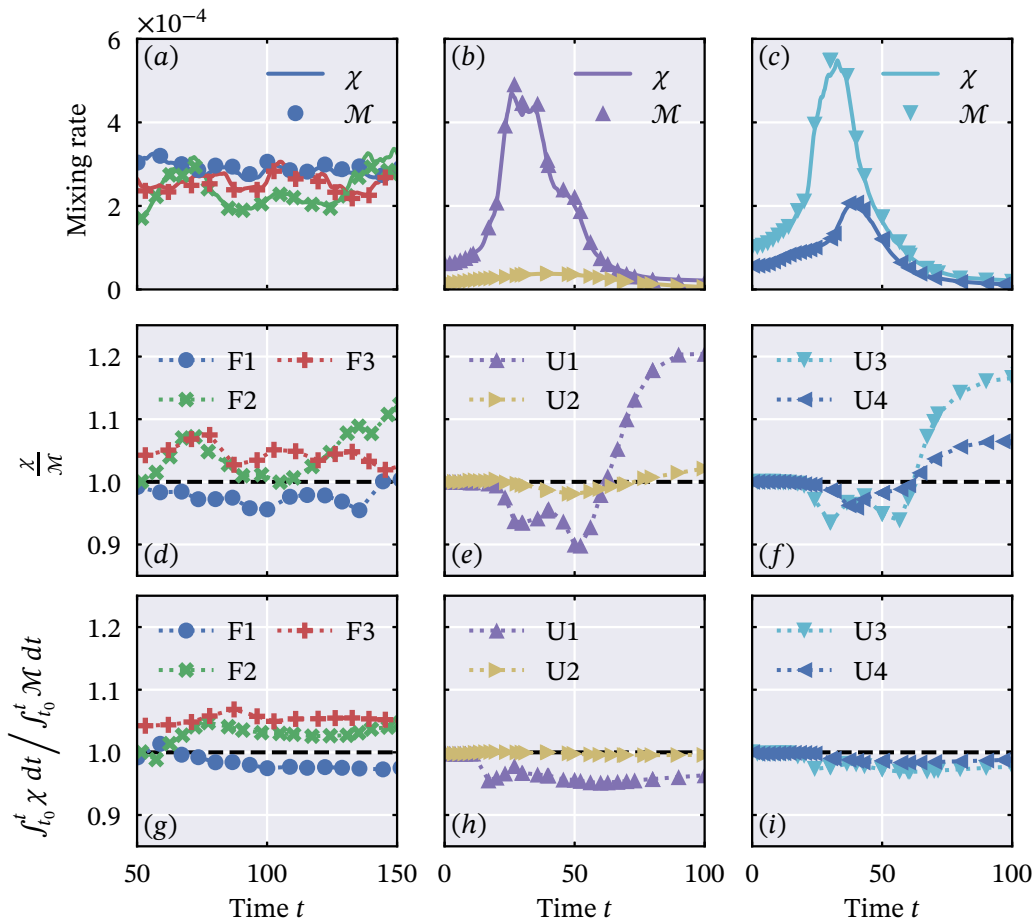}
    \caption{
        A time series comparison of the irreversible mixing rate $\mathcal{M}$ and the dissipation rate $\chi$ for each simulation in table \ref{tab:simulations}.
        $(a)$-$(c)$ plot the time series of $\mathcal{M}$ and $\chi$;
        $(d)$-$(f)$ plot the time series of $\chi/\mathcal{M}$ to highlight the fractional difference between the two;
        $(g)$-$(i)$ plot the time series of the ratio of the time-integrated quantities.
    }
    \label{fig:mix_chi}
\end{figure}

Nevertheless the dissipation rate $\chi$ is significantly more straightforward to quantify than the true diapycnal mixing rate $\mathcal{M}$, and so it is useful to investigate how well it can actually approximate the mixing.
The accuracy of $\chi$ for estimating mixing is also important in the context of ocean microstructure measurements, where small-scale gradients are measured directly but there is no way to obtain the relevant reference profile $b_*$.
In figures \ref{fig:mix_chi}a-c, we therefore plot the time series of both $\chi$ and $\mathcal{M}$ for each of our simulations.
By inspection, the two quantities appear to match up very well, with the symbols marking the mixing rate overlapping the lines plotting the time series of $\chi$.
To quantify how well $\chi$ approximates $\mathcal{M}$, we plot the time series of their ratio in figures \ref{fig:mix_chi}d-f.
Throughout the forced simulations, and for the early times of the unforced simulations, $\chi$ remains within 10\% of the true mixing rate.
At late times in simulations U1 and U3 the difference increases up to 20\%, but at this stage the flow is relaminarizing and $\mathcal{M}$ and $\chi$ are both small.
Indeed we show that this discrepancy is unimportant for quantifying the total mixing achieved over the course of the simulations in figures \ref{fig:mix_chi}g-i, where we plot the ratio of time-integrated $\chi$ and $\mathcal{M}$.
The time integral of $\mathcal{M}$ is equal to the increase in background potential energy due to diapycnal mixing, and we see that using $\chi$ to estimate this quantity results in at most a 5\% error in the total BPE change (corresponding to the final values of the cumulative ratio plotted in figures \ref{fig:mix_chi}g-i).

In the unforced simulations, $\chi$ consistently provides a slight underestimate of the diapycnal mixing rate.
This suggests that regions of intense turbulent mixing, associated with high values of $|\del b|$, preferentially sample regions where $\pd Z_\ast/\pd b>1$.
These regions are in turn associated with the reference buoyancy profile $b_*$ having a locally \emph{weaker} stratification than the mean.
In simulations F2 and F3, where forcing is applied in the form of internal gravity waves, the opposite is true and $\chi$ provides a slight overestimate of $\mathcal{M}$.
However it is not true that intense mixing occurs only in regions of strong or weak local stratification in each flow.
In all of the forced simulations for example, the standard deviation of $\pd Z_*/\pd b$ rises from the range 0.1-0.15 at time $t=50$ up to 0.25-0.3 at $t=150$, suggesting that as mixing persists throughout the simulations, the background profile is modified.
The fractional error between $\chi$ and $\mathcal{M}$ seen in figure \ref{fig:mix_chi}d does not show this increasing trend, suggesting that some local overestimates of $\mathcal{M}$ (where $\pd Z_*/\pd b<1$) cancel with some local underestimates (where $\pd Z_*/\pd b>1$) in the global average.
Similarly, the standard deviation of $\pd Z_*/\pd b$ reaches values in the range 0.15-0.2 for simulations U1 and U3 when $t>30$, approximately double the fractional error during the period of peak mixing.

\subsection{The effect of mean flow dissipation} \label{sec:mean_dissipation}

\begin{figure}
    \centering
    \includegraphics[width=\textwidth]{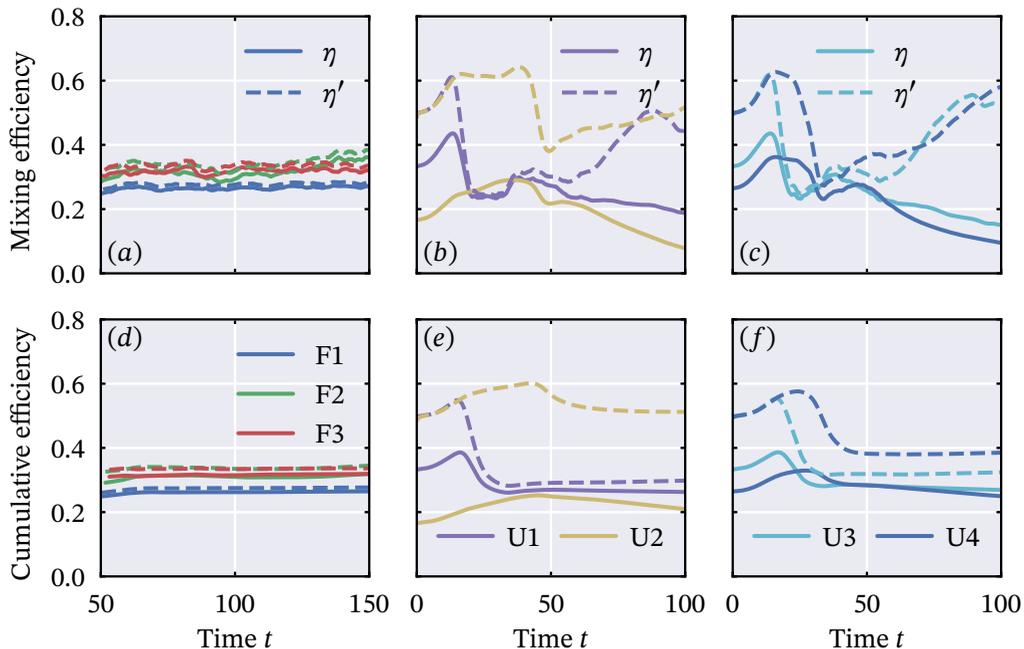}
    \caption{
        Time series of $(a)$-$(c)$ instantaneous and $(d)$-$(f)$ cumulative mixing efficiency, calculated with and without the mean flow dissipation, as defined in (\ref{eq:eta_prime_def}) and (\ref{eq:eta_c_def}) respectively.
    }
    \label{fig:mix_eff_mean}
\end{figure}

In the unforced simulations of set U, the majority of the kinetic energy is associated with the initial mean shear profile $\overline{u} = \sin z$.
At late times in these scenarios, the flow begins to relaminarize and the kinetic energy dissipation rate $\eps$ is dominated by the laminar diffusion of the mean shear.
Mixing efficiency is however often calculated using the \emph{turbulent} kinetic energy dissipation rate, that we quantify here as
\begin{equation}
    \eps^\prime = \frac{1}{Re}\left\langle \frac{\pd {u_i}^\prime}{\pd x_j}\frac{\pd {u_i}^\prime}{\pd x_j} \right\rangle ,
\end{equation}
where $\bu^\prime = \bu -\overline{\bu}$ is the velocity field perturbation from the horizontal average.
Figure \ref{fig:mix_eff_mean} compares time series of the following definitions of mixing efficiency calculated using either the turbulent dissipation rate $\eps^\prime$ or the total rate $\eps$
\begin{align}
    \eta &= \frac{\chi}{\chi+\eps} ,&
    \eta^\prime &= \frac{\chi}{\chi+\eps^\prime} .
    \label{eq:eta_prime_def}
\end{align}
We use $\chi$ rather than $\mathcal{M}$ in our definition of efficiency, since we have seen that the difference between them is small in the previous section, and our records of $\chi$ have better resolution in time.
Large discrepancies between $\eta$ and $\eta^\prime$ are observed when the average TKE dissipation rate $\eps^\prime$ is small compared to the dissipation rate of the mean flow $\overline{\eps}=\langle |\pd \overline{\bu}/\pd z|^2\rangle/Re$.
In simulation U2, wave breaking occurs at $t\approx 50$ and consists of small, strongly localised overturns that dissipate relatively quickly.
Consequentially $\eps^\prime$ remains smaller than $\overline{\eps}$ for the entire duration, leading to large differences between the efficiencies in figure \ref{fig:mix_eff_mean}b.
$\eta^\prime$ takes much larger values than $\eta$ in all of the unforced simulations at early and late times, with $\eta^\prime$ close to its initial value of 0.5.
This value corresponds to the diffusion associated with the plane wave form of \eqref{eq:wave_IC} and is a consequence of the choice $Pr=1$, that is molecular diffusion of buoyancy occurs at the same rate as the diffusion of momentum.
At larger values of $Pr$, diffusion of the wave would result in a far lower value of $\eta^\prime$.

\begin{table}
    \centering
    \label{tab:Reb_values}
    \begin{tabular}{l *{4}{S}}
        \toprule
        Simulation & \textrm{U1} & \textrm{U2} & \textrm{U3} & \textrm{U4} \\
        \midrule
        Maximum $Re_b(t)=\displaystyle\frac{\eps^\prime Re}{Ri_0}$ & 11.29 & 0.46 & 6.89 & 2.41 \\
        \midrule
        Maximum $Re_b(z) = \displaystyle\frac{\overline{\pd_j u_i^\prime \pd_j u_i^\prime}}{Ri_0 \left(1+\pd_z\overline{\theta}\right)}$ & 69.61 & 3.73 & 39.37 & 21.12 \\
        \bottomrule
    \end{tabular}
    \caption{
        Peak values of the buoyancy Reynolds number in the unforced simulations.
        The top row displays maximum (over $t$) values computed from the volume average $\eps^\prime$.
        The bottom row shows the maximum (over $z$) value of $Re_b$ computed from horizontal averages at the time instant of peak $\eps^\prime$.
    }
\end{table}

In figure \ref{fig:mix_eff_mean}d-f we also plot associated cumulative mixing efficiencies, defined here in terms of appropriate integrals of $\chi$ and $\eps$ (or $\eps^\prime$):
\begin{align}
    \eta_c &= \frac{\int_{t_0}^t \chi(t^\prime) \,\mathrm{d}t^\prime}{\int_{t_0}^t \chi(t^\prime) + \eps(t^\prime) \,\mathrm{d}t^\prime}, &
    \eta_c^\prime &= \frac{\int_{t_0}^t \chi(t^\prime) \,\mathrm{d}t^\prime}{\int_{t_0}^t \chi(t^\prime) + \eps^\prime(t^\prime) \,\mathrm{d}t^\prime}, \label{eq:eta_c_def}
\end{align}
where $t_0=50$ for the forced cases, and $t_0=0$ for the unforced cases.
The time integrals represent the energy changes associated with the cumulative effects of $\chi$ and $\eps$.
Figures \ref{fig:mix_eff_mean}e and \ref{fig:mix_eff_mean}f show that the diffusion of the mean shear flow has a significant impact on the total cumulative efficiency in the unforced simulations.
To emphasise that this is primarily a Reynolds number effect, we have listed measures of the buoyancy Reynolds number for the unforced simulations in table \ref{tab:Reb_values}.
Since the flows considered are extremely inhomogeneous in the vertical (as seen in figures \ref{fig:localAPE}e and \ref{fig:localAPE}f) we have calculated $Re_b$ from both volume and horizontal averages.
In oceanographic flows, we expect molecular diffusion to be negligible compared to the turbulent dissipation rate for the vast majority of the internal wave spectrum.
This result therefore highlights the challenge of using direct numerical simulations, where $Re$ is inevitably limited by computational resources, to investigate ocean mixing processes.

\section{Discussion and conclusions} \label{sec:discussion}

In this study, we have highlighted how the APE framework of \cite{winters_available_1995} should be generalised in the triply-periodic system often used in numerical simulations of stratified turbulent flows.
In these systems it is important to constrain the buoyancy field, inferred from the periodic perturbation $\theta$, to lie in a prescribed range.
We can then construct an accurate background buoyancy profile $b_*$ that is consistent with the periodic nature of the system.
However, setting limits on the buoyancy values effectively means that the shape of the domain can change in time.
In the case where the limiting buoyancy value has a non-overturning isopycnal surface, we find that this introduces an extra potential energy term $\mathcal{S}$ as defined in \eqref{eq:S_def}.
Appropriate definitions of available and background potential energy can then be obtained by accounting for this additional term as in \eqref{eq:B_def} and \eqref{eq:A_def}.

Constructing the correct background profile is also vital for accurately calculating the local APE density $E_{APE}$ defined by \cite{scotti_diagnosing_2014}.
This quantity can then provide useful information for identifying mechanisms by which mixing can occur.
When integrated over the domain, the local APE also recovers all of the additional terms in our new global APE variable $\mathcal{A}$.
Furthermore, the local APE can even be quantified in scenarios where our global APE is not well defined.
So long as the background profile $b_*$ is identified, both $E_A\equiv\langle E_{APE} \rangle$ and the irreversible, diapycnal mixing rate $\mathcal{M}$ can be calculated.
The evolution of $E_A$ is then entirely determined by the mixing rate and the buoyancy flux, with zero contribution from boundary fluxes.
We can therefore calculate the exact rate of diapycnal mixing in more energetic stratified flows that use periodic domains, such as those considered by \cite{de_bruyn_kops_effects_2019} and \cite{portwood_asymptotic_2019}.

This technique for calculating APE could also be applied to \emph{unstably} stratified periodic systems, where $Ri_0 < 0$, used to study bulk properties of convection \citep[e.g.][]{lohse_ultimate_2003}.
In traditional Rayleigh--B\'enard convection, \cite{gayen_completing_2013} find that irreversible mixing is largely confined to thermal boundary layers.
It would therefore be interesting to investigate whether the theoretical prediction of $\eta\rightarrow 0.5$ at high $Ra$ holds for the periodic convection setup, where such boundary layers are absent.
The sorting technique presented here is also applicable to the case of passive scalar flows where a mean gradient is imposed.
Such setups are useful for studying the mixing of biogeochemical tracers, which are often found with significant mean gradients in the ocean \citep{williams_ocean_2011}.
Although the concept of APE would not be relevant here, sorting between isoscalar surfaces would provide an appropriate background profile to enable accurate calculation of the diascalar flux as proposed by \citet{winters_diascalar_1996}.

\begin{figure}
    \centering
    \includegraphics[width=\textwidth]{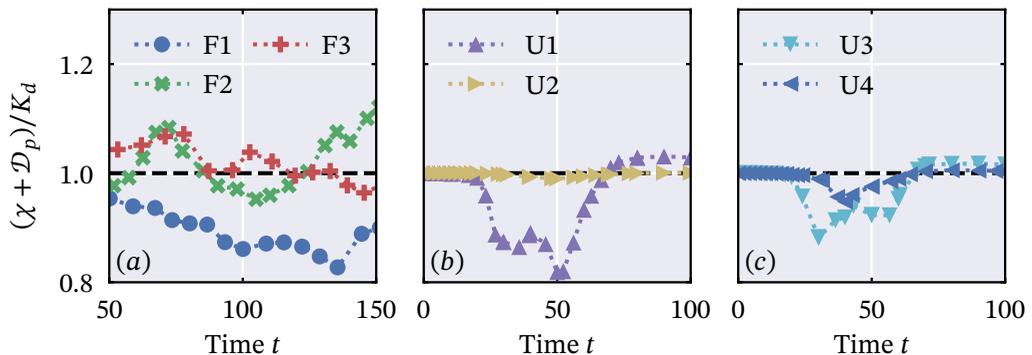}
    \caption{
        A time series comparison of the diapycnal diffusivity $K_d$ and the approximation of $\chi+\mathcal{D}_p$.
        The ratio of the two is plotted in an analogous fashion to figures \ref{fig:mix_chi}d-f.
    }
    \label{fig:diffusivity}
\end{figure}

In observational oceanography, turbulent mixing can be estimated by using fast-response thermistors to measure small-scale temperature gradients.
The primary aim in this context is to estimate a diapycnal diffusivity, defined in our dimensionless formulation as
\begin{equation}
    K_d = \frac{Ri_0}{RePr} \left\langle \left(\left.\frac{\pd Z_*}{\pd b}\right|_{b(\bx,t)}\right)^2 \left|\del b\right|^2 \right\rangle = \frac{Ri_0}{RePr} \left\langle \frac{\left| \del b\right|^2}{\left(\left.\pd b_*/\pd z\right|_{z_*(\bx,t)}\right)^2} \right\rangle . \label{eq:diapycnal_diffusivity}
\end{equation}
Since the reference profile $b_*$ cannot be obtained in the ocean, a large-scale average is taken of the buoyancy (or temperature) gradient.
The estimate often attributed to \cite{osborn_oceanic_1972} is then used such that
\begin{equation}
    K_d \approx \frac{Ri_0}{RePr} \frac{\left\langle \left| \del \theta \right|^2 \right\rangle}{\left\langle \pd b/\pd z \right\rangle^2} =\chi. \label{eq:osborn-cox}
\end{equation}
Note that the internal energy conversion rate $\mathcal{D_p}$ is neglected here, since it is assumed to be much smaller than $\chi$ in a turbulent flow.
In dimensional form it is common to see \eqref{eq:osborn-cox} written as $K_d=\chi/N^2$, but in our non-dimensionalization the mean buoyancy gradient in the denominator is prescribed to be equal to one.
The approximation made in estimating $K_d$ in \eqref{eq:osborn-cox} is the same approximation used in \S\ref{sec:chi_estimate} to estimate the mixing rate $\mathcal{M}$ with $\chi$.
Precisely, we approximate the reference buoyancy gradient $\pd b_\ast/\pd z$ by the imposed mean stratification.
We test this approximation in the context of diapycnal diffusivity in figure \ref{fig:diffusivity} by plotting the time series of $(\chi+\mathcal{D}_p)/K_d$.
The fractional error between the estimate $\chi+\mathcal{D}_p$ and the true diffusivity remains within one standard deviation of $\pd Z_\ast/\pd b$ for every simulation.
Figures \ref{fig:diffusivity}b and \ref{fig:diffusivity}c show that $\chi+\mathcal{D}_p$ underestimates the diffusivity at the time of most intense mixing in the unforced simulations.
This reaffirms the conclusion drawn from figures \ref{fig:mix_chi}e and \ref{fig:mix_chi}f that the turbulent mixing in this flow preferentially samples regions of relatively weak local stratification.
\cite{salehipour_diapycnal_2015} find a similar underestimation of $K_d$ in turbulent flows developing from Kelvin--Helmholtz instability in a stratified shear layer.
An investigation to identify in which flows \eqref{eq:osborn-cox} provides an over/underestimate of the diffusivity would be valuable for understanding the variability associated with the approximation.

We include the internal energy conversion rate $\mathcal{D}_p$ in our estimate in figure \ref{fig:diffusivity} since it is not always negligible in the simulations.
Furthermore \cite{gregg_mixing_2018} remark that $\mathcal{D}_p$ should be included when applying mixing results to the strongly stratified pycnocline where mixing is localised and intermittent.
In the periodic setup studied here, the boundary flux $\mathcal{F}_d$ counteracts $\mathcal{D}_p$ in the BPE energy budget \eqref{eq:B_evo} to maintain the constant mean stratification.
When quantifying diffusivity in this system, it is therefore important to include the contribution from $\mathcal{D}_p$ and to compute $\mathcal{M}+\mathcal{D}_p$ directly, instead of relying on changes in BPE.
In many observational studies focused on mixing in turbulent patches (where $\mathcal{D}_p$ is negligible), practical difficulties in obtaining an accurate value of $\chi$ result in far larger implied levels of uncertainty than those apparent in in figure \ref{fig:diffusivity}
\citep[see for example][]{waterhouse_global_2014}.
In this sense our results show that \eqref{eq:osborn-cox} provides a good estimate of the diapycnal diffusivity in the stratified flows considered.

In the case of \emph{homogeneous} turbulence subject to a uniform mean stratification, \citet{stretch_diapycnal_2010} show $\chi$ and $\mathcal{M}$ to be equivalent.
Indeed, if such homogeneous turbulence is maintained in a steady state by energy transfers from the velocity field, then $\chi$ is also equivalent to $-\J$.
This is exactly the reasoning of \citet{osborn_oceanic_1972}.
In the periodic system considered here, if mixing were homogeneous throughout the domain then the boundary flux $\mathcal{F}_d$ would also balance the interior diapycnal flux $\mathcal{M}+\mathcal{D}_p$ such that the BPE equation \eqref{eq:B_evo} was steady.
As highlighted by \citet{portwood_robust_2016}, many stratified flows are \emph{not} homogeneous in this sense, with turbulence becoming more patchy and intermittent when subject to stronger stratification.
This even applies to flows where the initial state is homogeneous, such as the decay of a turbulent cloud in an initially uniform stratification \citep[e.g.][]{bartello_sensitivity_2013}.
We believe the APE framework presented above will prove useful in determining the potential impacts of such developing inhomogeneities.

Due to the aforementioned difficulties involved in accurately resolving small-scale temperature gradients, shear probes are used more frequently than thermistors to infer mixing rates in the ocean.
Further assumptions are however needed to obtain mixing estimates from such velocity gradient measurements.
On top of the \cite{osborn_oceanic_1972} model, the buoyancy variance destruction rate may be approximated by $\chi \simeq -\mathcal{J}=\Gamma \eps$, where the turbulent flux coefficient $\Gamma$ is taken to be a constant, usually 0.2 in practice after \cite{osborn_estimates_1980}, under a set of assumptions that the turbulent flow is, for example quasi-steady.
The turbulent flux coefficient is related to the mixing efficiency defined in \eqref{eq:eta_prime_def} through
\begin{equation}
    \Gamma = \frac{\eta}{1-\eta} . \label{eq:Gamma}
\end{equation}
Many experimental and numerical studies have shown variation in the mixing efficiency across a range of stratified flows, as reviewed by \cite{ivey_density_2008} and \cite{caulfield_layering_2021}.
This has motivated a body of work to investigate the functional dependence of $\eta$ on various dimensionless parameters, including the Richardson number, buoyancy Reynolds number $Re_b=\eps/\nu N^2$, and turbulent Froude number $Fr=\eps/N\K$.
Despite this concerted effort to provide insight into how $\eta$ varies, there is no clear physical explanation as to why $\Gamma=0.2$ is a sensible assumption or why it appears to provide diffusivity estimates in line with those from tracer release experiments \citep{ledwell_evidence_2000}.
In figure \ref{fig:mix_eff_mean} we highlight examples where laminar diffusion of a shear flow can strongly impact the calculated values of $\eta$.
Although not relevant for high Reynolds number flows found in the ocean, it is important to acknowledge the effect of this diffusion in idealised numerical studies that discuss mixing efficiency in the context of ocean mixing.
This is most relevant for flows where turbulence is transient and localised, such as those arising from instabilities in stratified shear layers.

In the context of estimating mixing from oceanic measurements, the poorly constrained variability of $\Gamma$ implies that the \citet{osborn_oceanic_1972} model will instead provide the best estimates of mixing from microstructure data.
Indeed our results above show that in turbulent flows with a large-scale mean buoyancy gradient $N_0^2$, the \citet{osborn_oceanic_1972} model \eqref{eq:osborn-cox} provides a reliable estimate of the diapycnal diffusivity.
In oceanic flows calculating $N_0$ can however prove challenging, particularly in the case of internal waves breaking close to boundaries as highlighted by \citet{arthur_how_2017}.
Given the significant dissipation of energy close to boundaries in the ocean, the calculation of $N_0$ in such flows remains an important outstanding issue for estimating diapycnal mixing in these regions.
Furthermore, the \citet{osborn_oceanic_1972} method can only be applied in regions where the ocean is stratified by temperature.
Where salt acts as a stratifying agent, the turbulent flux coefficient $\Gamma$ must be specified to obtain microstructure mixing estimates through the \citet{osborn_estimates_1980} method.

In particular, for the energetic framework presented here to be truly applicable to real oceanographic flows, there are at least three open issues which need to be addressed.
First, it is not at all clear what the effect of more realistic Reynolds numbers, or indeed realistically higher values of $Pr = O(10-1000)$ will have on the various mixing properties and energetic pathways discussed here.
Second, it is still an open question of some controversy whether ${\Gamma \approx 0.2}$, or equivalently $\eta \approx 1/6$, is actually `typical' of quasi-steady mixing processes, or whether $\Gamma$ actually depends on parameters of the flow.
\cite{portwood_asymptotic_2019} recently demonstrated the emergence of ${\Gamma=0.2}$ in sheared DNS that was controlled by construction to be quasi-steady.
It is at least plausible that the higher values of efficiency observed for the flows discussed here are artefacts of the inherent transience of these flows.
Of course, mixing events in the ocean are likely to be highly spatio-temporally intermittent, not least because of the key role played by `breaking' internal waves, as argued by \cite{mackinnon_climate_2017} and modelled here, so the relevance of quasi-steady sustained stratified turbulence to the real ocean is not immediately obvious.
Thirdly, complications associated with layered states, either due to hydrodynamic mechanisms associated with turbulence \citep{caulfield_layering_2021} or associated with double-diffusive convection \citep{schmitt_double_1994} are clearly of interest.
The energetic framework presented here is nevertheless well-suited to address these three open issues, or indeed other challenges of real relevance to the quantification and parameterization of mixing in realistic stratified flows.

\section*{Acknowledgements}
We are grateful to R.\ Tailleux and two anonymous referees for their insightful and constructive comments on this manuscript.
The work of C.J.H.\ was supported by the Cambridge Earth System Science NERC DTP (NE/L002507/1).
The data and Python notebooks used to create the figures, as well as a Python script implementing the sorting method described above, are freely available at \url{https://doi.org/10.17863/CAM.58957}.

\section*{Declaration of interests}
The authors report no conflict of interest.

\appendix
\section{Consideration of a more general boundary isopycnal} \label{app:iso}

In \eqref{eq:z1_def}, we assume that the boundary buoyancy contour can be parameterized by $x$ and $y$.
Now let us consider a more general isopycnal boundary that may overturn, where the surface of constant buoyancy is parameterized by arbitrary coordinates $p$ and $q$.
The implicit definition of the isopycnal surface $\bx_1(p,q,t)$ is then given by
\begin{equation}
    b( x_1(p,q,t), y_1(p,q,t), z_1(p,q,t), t) = b_0 . \label{eq:iso_gen}
\end{equation}
Considering the same volume integral as in \eqref{eq:Leibniz}, we apply the Reynolds Transport Theorem to obtain
\begin{equation}
    \frac{d}{dt}\left(\int_V f \dV\right) = \int_V \frac{\pd f}{\pd t} \dV + \int_S \left(f|_{\bx=\bx_2} - f|_{\bx=\bx_1}\right) \frac{\pd \bx_1}{\pd t} \cdot \symbfit{n} \, \mathrm{d}S. \label{eq:RTT} 
\end{equation}
$S$ denotes the domain in $(p,q)$ space that parameterizes the surface, $\bx_1 = (x_1, y_1, z_1)$ is the location of the isopycnal surface in Cartesian coordinates, and the area element is given by
\begin{equation}
    \symbfit{n} \,\mathrm{d}S = \left(\frac{\pd \bx}{\pd p} \times \frac{\pd \bx}{\pd q}\right) \,\mathrm{d}p\,\mathrm{d}q .
    \label{eq:normal_arbitrary}
\end{equation}
Note that for $p=x$ and $q=y$, this recovers the original Leibniz rule result of \eqref{eq:Leibniz} since $\bx_1 = (x, y, z_1(x,y,t))$ and
\begin{equation}
    \symbfit{n} \,\mathrm{d}S = \frac{\del b}{\pd b/\pd z} \,\mathrm{d}x \,\mathrm{d}y . \label{eq:normal_z1}
\end{equation}
We know in general that the direction of the normal is that of the buoyancy gradient $\del b$, but for the arbitrary form \eqref{eq:normal_arbitrary} the magnitude of $\bx_p\times\bx_q$ depends on the coordinates chosen.
Since we wish to calculate the surface integral from simulation data, it is convenient to restrict ourselves to non-overturning isopycnals, where the magnitude of the area element can be straightforwardly obtained.

We can however manipulate \eqref{eq:RTT} further by noting that $\symbfit{n}=\del b/|\del b|$, and defining the average over the surface $S$ as
\begin{equation}
    \overline{f}^* = \frac{1}{A_S}\int_S f \,\mathrm{d}S ,
\end{equation}
where $A_S$ is the surface area of the isopycnal defined in \eqref{eq:iso_gen}.
Applying this to the Reynolds transport theorem result \eqref{eq:RTT} gives
\begin{equation}
    \frac{d}{dt}\left(\int_V f \dV\right) = \int_V \frac{\pd f}{\pd t} \dV + \overline{\left[f\right]_{b=b_0}^{b=b_0+L_z} \frac{\pd \bx_1}{\pd t} \cdot \frac{\del b}{|\del b|}}^* .
\end{equation}
Substituting $f=-Ri_0bz$ to find the extra term in the potential energy equation provides
\begin{equation}
    \frac{d\P}{dt} = -Ri_0 \left\langle z\frac{\pd b}{\pd t}\right\rangle - Ri_0 \frac{A_S}{A} \overline{(b_0+z_2)\frac{\pd \bx_1}{\pd t}\cdot \frac{\del b}{|\del b|}}^*, \label{eq:P_RTT}
\end{equation}
where $A$ is the cross-sectional area of the domain in the $x$-$y$ plane, and from \cite{winters_diascalar_1996} we know that
\begin{equation}
    \frac{A_S}{A} = \frac{\pd Z_*}{\pd b} \frac{\overline{|\del b|^2}^*}{\overline{|\del b|}^*}.
\end{equation}
Although the last term in \eqref{eq:P_RTT} can be expressed analytically, its computation is far more arduous than $-d\mathcal{S}/dt$, and it does not appear (thus far) to simplify to a similar form.

\section{Derivation of the potential energy equations}

\subsection{Total potential energy} \label{app:tot_PE}
In this section, the control volume is bounded by the isopycnals $b=b_0$ ($z_1$) and $b=b_0+L_z$ ($z_2$).
Consider the time evolution of $\P=-Ri_0 \langle bz\rangle$ by applying the Leibniz rule as in \eqref{eq:Leibniz}:
\begin{align}
    \frac{d\P}{dt} &= -Ri_0 \left\langle \frac{\pd (bz)}{\pd t} \right\rangle - \frac{Ri_0}{V}\int_A \left[ bz \right]_{z=z_1}^{z_2} \frac{\pd z_1}{\pd t} \dA , \\
    &= -Ri_0 \left\langle z\frac{\pd b}{\pd t} \right\rangle - \frac{Ri_0}{V} \int_A L_z(b_0 + z_2) \frac{\pd z_2}{\pd t} \dA, \\
    &= -Ri_0 \left\langle z\frac{\pd b}{\pd t} \right\rangle - Ri_0 b_0 \frac{d\overline{z_2}}{dt} - Ri_0 \frac{d}{dt} \left( \frac{\overline{{z_2}^2}}{2}\right).
\end{align}
Defining $\mathcal{S}$ as in \eqref{eq:S_def}, we move the last two terms in the above equation to the right hand side, and use the buoyancy evolution equation \eqref{eq:NS3} to expand the first term as
\begin{align}
    \frac{d\P}{dt} + \frac{d\mathcal{S}}{dt} &= -Ri_0\left\langle z\left( -\bu\cdot\del b + \frac{1}{RePr}\nabla^2 b\right) \right\rangle, \\
    &= Ri_0 \left\langle \del \cdot (zb\bu ) - wb - \frac{z}{RePr} \del \cdot \del b \right\rangle, \\
    &= Ri_0 \left\langle \del \cdot (zb\bu)\right\rangle - Ri_0\left\langle w\theta\right\rangle - Ri_0 \left\langle wz \right\rangle - \frac{Ri_0}{RePr}\left\langle \del \cdot\left(z\del b\right) - \del z \cdot\del b\right\rangle, \\
    &= Ri_0 \left\langle \del \cdot (zb\bu)\right\rangle - \mathcal{J} - Ri_0 \left\langle \del \cdot \left(\frac{z^2}{2}\bu \right) \right\rangle - \frac{Ri_0}{RePr} \left\langle \del \cdot\left(z\del b\right)\right\rangle + \mathcal{D}_p.
\end{align}
The final term in the above equation is obtained through
\begin{equation}
    \frac{Ri_0}{RePr} \left\langle \del z \cdot \del b \right\rangle = \frac{Ri_0}{RePr} \left\langle \frac{\pd b}{\pd z} \right\rangle = \frac{Ri_0}{VRePr} \int_A \left[ b \right]_{z=z_1}^{z=z_2} \dA = \frac{Ri_0}{RePr} \equiv \mathcal{D}_p
\end{equation}
With the boundaries we have specified, the divergence theorem for an arbitrary vector field $\symbfit{f}(\bx,t)$ takes the form
\begin{equation}
    \int_V \del \cdot \symbfit{f} \dV = \int_A \left[ \symbfit{f} \right]_{z=z_1}^{z_2} \cdot \frac{\del b}{\pd b/\pd z} \dA , \label{eq:div_thm}
\end{equation}
where $\del b/(\pd b/\pd z)$ is evaluated on the surface $z=z_1$ (and takes the same value on the surface $z=z_2$).
Applying the divergence theorem to each of the above terms then gives
\begin{align}
    \left\langle \del\cdot\left(zb\bu\right)\right\rangle &= \frac{1}{V}\int_A \left[zb\bu\right]_{z_1}^{z_2} \cdot \frac{\del b}{\pd b/\pd z} \dA = \frac{1}{A}\int_A \left(b_0+z_2\right) \left[\frac{\bu\cdot\del b}{\pd b/\pd z}\right]_{z_1} \dA , \\
    \left\langle \del\cdot\left(z^2\bu/2\right)\right\rangle &= \frac{1}{V}\int_A \left[\frac{z^2\bu}{2}\right]_{z_1}^{z_2} \cdot \frac{\del b}{\pd b/\pd z} \dA = \frac{1}{A}\int_A \left(\frac{L_z}{2}+z_1\right) \left[\frac{\bu\cdot\del b}{\pd b/\pd z}\right]_{z_1} \dA , \\
    \left\langle\del\cdot\left(z\del b\right)\right\rangle &= \frac{1}{V}\int_A \left[z\del b\right]_{z_1}^{z_2} \cdot \frac{\del b}{\pd b/\pd z} \dA = \frac{1}{A}\int_A \left[\frac{|\del b|^2}{\pd b/\pd z}\right]_{z_1} \dA .
\end{align}
The potential energy evolution therefore simplifies to
\begin{equation}
    \frac{d\P}{dt} + \frac{d\mathcal{S}}{dt} = -\mathcal{J} - \mathcal{F}_d + \mathcal{D}_p + \left(b_0+\frac{L_z}{2}\right) \frac{1}{A}\int_A \left[\frac{\bu\cdot\del b}{\pd b/\pd z}\right]_{z_1} \dA.
\end{equation}
We can show that this final integral is zero by considering the evolution of the volume-averaged buoyancy.
Since $b=z+\theta$, we know that $\langle b\rangle = L_z/2 + \overline{z_1}+\langle \theta\rangle$.
The mean buoyancy perturbation is coupled to the mean vertical velocity through the system
\begin{align}
    \frac{d\langle\theta\rangle}{dt} &= \left\langle \frac{\pd \theta}{\pd t} \right\rangle = - \langle w\rangle, &
    \frac{d\langle w\rangle}{dt} &= \left\langle \frac{\pd w}{\pd t}\right\rangle = Ri_0 \langle\theta\rangle.
\end{align}
Importantly, if both $\langle\theta\rangle$ and $\langle w\rangle$ are initially zero, then they remain so forever.
This is the scenario most commonly applied in studies using the periodic stratified setup, so we proceed taking $\langle\theta\rangle\equiv 0$.
We therefore know that
\begin{equation}
    \frac{d\langle b\rangle}{dt} = \frac{d\overline{z_1}}{dt} .
\end{equation}
Applying the Leibniz rule of \eqref{eq:Leibniz} to $\langle b\rangle$ instead gives
\begin{equation}
    \frac{d\langle b\rangle}{dt} = \left\langle \frac{\pd b}{\pd t} \right\rangle + \frac{1}{V}\int_A\left[b\right]_{z_1}^{z_2} \frac{\pd z_1}{\pd t} \dA = \left\langle \frac{\pd b}{\pd t} \right\rangle + \frac{d\overline{z_1}}{dt} .
\end{equation}
We can then deduce that the desired integral is zero as follows
\begin{align}
    0 &= \left\langle \frac{\pd b}{\pd t}\right\rangle
    = \left\langle -\del\cdot \left(b\bu\right) + \frac{1}{RePr} \del \cdot \del b \right\rangle, \\
    &= -\frac{1}{V} \int_A \left[ b\bu\right]_{z_1}^{z_2} \cdot \frac{\del b}{\pd b/\pd z} \dA + \frac{1}{VRePr}\int_A \left[ \del b\right]_{z_1}^{z_2} \cdot \frac{\del b}{\pd b/\pd z} \dA, \\
    &= -\frac{1}{A}\int_A \left[\frac{\bu\cdot\del b}{\pd b/\pd z}\right]_{z_1} \dA ,
\end{align}
where we have applied the divergence theorem and used that $\del b|_{z_1}=\del b|_{z_2}$.

\subsection{Background potential energy} \label{app:BPE}

In this section, we set $b_0=0$ so the boundary surfaces $z_1$ and $z_2$ correspond to the isopycnals $b=0$ and $b=L_z$.
We begin by determining the time evolution of $\P_B = -Ri_0\langle b z_* \rangle$.
Applying the Leibniz result of \eqref{eq:Leibniz} to this quantity gives
\begin{align}
    \frac{d\P_B}{dt} &= -Ri_0 \left\langle \frac{\pd (b z_*)}{\pd t} \right\rangle - \frac{Ri_0}{V} \int_A \left[ bz_* \right]_{z=z_1}^{z_2} \frac{\pd z_1}{\pd t} \dA ,\\
    &= -Ri_0 \left\langle z_*\frac{\pd b}{\pd t} + b\frac{\pd z_*}{\pd t} \right\rangle - \frac{Ri_0}{V} \int_A L_z \overline{z_2} \frac{\pd z_1}{\pd t} \dA ,\\
    &= -Ri_0 \left\langle z_*\frac{\pd b}{\pd t}\right\rangle - Ri_0 \left\langle b\frac{\pd z_*}{\pd t} \right\rangle - Ri_0 \overline{z_2} \frac{d \overline{z_2}}{dt} . \label{eq:PB_working}
\end{align}
The second term in the line above is zero in the case of fixed, insulating, horizontal boundaries.
We therefore consider the simple case of $\theta=-z_1(x,y,t)$ to investigate the contribution this term has in the case of time-dependent isopycnal boundaries.
As in \S\ref{sec:localAPE}, this example has the linear background profiles $Z_*(s,t)=s+\overline{z_1}$, and $b_*(s,t) = s - \overline{z_1}$, so
\begin{equation}
    z_*(\bx,t) = Z_*(b(\bx,t), t)=b(\bx,t) + \overline{z_1}(t)=z-z_1(x,y,t) + \overline{z_1}(t) . \label{eq:example_field}
\end{equation}
For this simple example we find that
\begin{equation}
    \left\langle b \frac{\pd z_*}{\pd t} \right\rangle = 0 , \label{eq:Zt_example}
\end{equation}
and conclude that there is no additional contribution to this term when considering a moving boundary.
We now consider the first term in \eqref{eq:PB_working}, and use the buoyancy evolution equation \eqref{eq:NS3} to obtain
\begin{align}
    \left\langle z_* \frac{\pd b}{\pd t} \right\rangle &= \left\langle z_* \left( -\bu\cdot\del b + \frac{1}{RePr} \nabla^2 b \right) \right\rangle, \\
    &= \left\langle -\bu\cdot z_*\del b + \frac{1}{RePr} z_*\del \cdot \del b \right\rangle, \\
    &= \left\langle -\del\cdot(\psi\bu) + \frac{1}{RePr}\left(\del\cdot(z_*\del b) - \del z_* \cdot \del b \right) \right\rangle. \label{eq:z*dbdt}
\end{align}
Here we have introduced the Casimir
\begin{equation}
    \psi(b)=\int_0^b Z_*(s) \,\mathrm{d}s ,
\end{equation}
that satisfies $\del\psi=z_*\del b$.
Since $Z_*$ is the inverse of $b_*$, and we know that $\langle b_*(z)\rangle = \langle b(\bx) \rangle$, we can furthermore deduce that
\begin{equation}
    \psi(L_z) = \int_0^{L_z} Z_*(s) \,\mathrm{d}s = L_z \overline{z_2} - \int_{\overline{z_1}}^{\overline{z_2}} b_*(s) \mathrm{d}s = \frac{{L_z}^2}{2} . \label{eq:Casimir_int}
\end{equation}
We also note that $\del z_* = (\pd Z_*/\pd b)\del b$, and this can be applied to the final term in \eqref{eq:z*dbdt}.
Applying the divergence theorem \eqref{eq:div_thm} to the term involving the Casimir produces
\begin{equation}
    \left\langle \del \cdot (\psi \bu)\right\rangle = \frac{L_z}{2A}\int_A \left[\frac{\bu\cdot\del b}{\pd b/\pd z}\right]_{z=z_1} \dA = 0.
\end{equation}
Only the diffusive terms remain, giving
\begin{align}
    -Ri_0 \left\langle z_* \frac{\pd b}{\pd t} \right\rangle &= \frac{-Ri_0}{RePr} \left( \frac{1}{A}\int_A \left[\frac{|\del b|^2}{\pd b/\pd z}\right]_{z=z_1} \dA - \left\langle \frac{\pd Z_*}{\pd b}|\del b|^2 \right\rangle \right) , \\
    &= -\mathcal{F}_d + \mathcal{M} + \mathcal{D}_p.
\end{align}
We now have
\begin{equation}
    \frac{d\mathcal{P}_B}{dt} = \mathcal{M} + \mathcal{D}_p - \mathcal{F}_d - Ri_0 \frac{d}{dt}\left( \frac{\overline{z_2}^{\,2}}{2} \right).
\end{equation}
Defining $\mathcal{B}=\mathcal{P}_B + Ri_0 \overline{z_2}^{\,2}/2$ as in \eqref{eq:B_def}, we finally arrive at the evolution equation
\begin{align}
    \frac{d\mathcal{B}}{dt} &= \mathcal{M} + \mathcal{D}_p - \mathcal{F}_d .
\end{align}

\section{Equivalence of various local APE definitions for an adiabatically sorted buoyancy profile} \label{app:APE_equivalence}

\citet{tailleux_available_2013-1} proposes the following APE density as work against buoyancy forces defined relative to an arbitrary $z$-dependent reference density profile $\rho_r(z,t)$:
\begin{equation}
    \mathcal{E}_a(S_i,T,z,t) = \int_{z_r}^z \frac{g}{\rho_0} \left(\rho(S_i, T, z^\prime) - \rho_r(z^\prime,t) \right) \,\mathrm{d}z^\prime. \label{eq:remi_APE}
\end{equation}
Here the density field depends on a materially conserved temperature variable $T$ as well as an arbitrary number of compositional variables $S_i$, and $z_r$ is the level of neutral buoyancy satisfying $\rho(S_i, T, z_r)=\rho_r(z_r,t)$.
The above expression generalises the `potential energy density' of \citet{andrews_note_1981} to an arbitrary nonlinear equation of state.
Although \eqref{eq:remi_APE} only applies under the Boussinesq approximation, this expression can be extended as in \citet{tailleux_local_2018} to describe APE density for a \emph{compressible} multicomponent fluid.
The arbitrary reference profile can be useful for defining alternative measures of APE.
For example if the uniform, mean gradient is taken as the reference buoyancy profile $b_r=z$, then \eqref{eq:remi_APE} recovers the APE defined in \eqref{eq:Lorenz_def}.

In this study, we consider a Boussinesq fluid with a linear equation of state in one variable, and take the reference profile to be the adiabatically sorted buoyancy $b_r=b_*$.
With these assumptions, and applying our non-dimensionalisation, \eqref{eq:remi_APE} becomes
\begin{equation}
    \mathcal{E}_a(\bx, t) = -Ri_0 \int_{z_*(\bx,t)}^z b(\bx, t) - b_*(z^\prime, t) \,\mathrm{d}z^\prime . \label{eq:roullet_APE}
\end{equation}
This expression is exactly that used by \citet{roullet_available_2009}.
Note that \eqref{eq:roullet_APE} can also be rearranged into the form
\begin{equation}
    \mathcal{E}_a(\bx, t) = -Ri_0 \left(z-z_*(\bx,t)\right) \left[b(\bx,t) - \frac{1}{z-z_*(\bx,t)} \int_{z_*(\bx,t)}^z b_*(z^\prime, t) \,\mathrm{d}z^\prime \right] , \label{eq:Barkan_APE}
\end{equation}
the expression for APE density used by \citet{winters_available_2013}.

We can further relate \eqref{eq:roullet_APE} to the definition of APE by \citet{scotti_diagnosing_2014} (which itself is equivalent to the original definition of \citet{holliday_potential_1981} but with simpler notation).
We rewrite \eqref{eq:roullet_APE} as
\begin{equation}
    \mathcal{E}_a = -Ri_0 b(z-z_*) + Ri_0 \int_{z_*}^z b_*(z^\prime, t) \,\mathrm{d}z^\prime , \label{eq:locAPE_step}
\end{equation}
and make the substitution $z^\prime = Z_*(s,t)$, where $Z_*$ is the inverse map of the sorted buoyancy profile $b_*$.
The integral part of \eqref{eq:locAPE_step} then becomes
\begin{equation}
    \int_{z_*}^z b_*(z^\prime, t) \,\mathrm{d}z^\prime = \int_b^{b_*} s \frac{\pd Z_*}{\pd s} \mathrm{d}s ,
\end{equation}
since $b_*(Z_*(s,t),t) = s$ and $b_*(z_*(\bx,t), t) = b_*(Z_*(b(\bx,t),t), t) = b(\bx,t)$.
Integrating by parts then leads to
\begin{align}
    \int_{z_*}^z b_*(z^\prime, t) \,\mathrm{d}z^\prime &= \left[sZ_*(s,t)\right]_{s=b}^{s=b_*} - \int_b^{b_*} Z_*(s,t) \,\mathrm{d}s ,\\
    &= b_* z - b z_* - \int_{b}^{b_*} Z_*(s,t) \,\mathrm{d}s .
\end{align}
Finally, we can substitute this expression into \eqref{eq:locAPE_step} to recover the form of \citet{holliday_potential_1981} and \citet{scotti_diagnosing_2014}:
\begin{align}
    \mathcal{E}_a &= -Ri_0 \left[bz - bz_* - b_* z + bz_* + \int_{b}^{b_*} Z_*(s,t) ds \right] , \\
    \Rightarrow \mathcal{E}_a &= E_{APE} \equiv -Ri_0 \int_{b_*}^b z - Z_*(s,t) \,\mathrm{d}s .
\end{align}

\bibliographystyle{jfm}

\bibliography{Special_Issue}

\end{document}